\documentclass[manuscript,screen]{acmart}

\AtBeginDocument{%
  }

\setcopyright{acmlicensed}
\copyrightyear{2024}
\acmYear{2024}
\acmDOI{XXXXXXX.XXXXXXX}

\acmISBN{978-1-4503-XXXX-X/18/06}

\author{Ruichao Liang}
\email{liangruichao@whu.edu.cn}
\affiliation{%
  \institution{Wuhan University}
  \city{Wuhan}
  \country{China}
}

\author{Jing Chen}
\email{chenjing@whu.edu.cn}

\affiliation{%
  \institution{Wuhan University}
  \city{Wuhan}
  \country{China}
}

\author{Cong Wu}
\email{cong.wu@ntu.edu.sg}
\affiliation{%
  \institution{Nanyang Technological University}
  \country{Singapore}
}

\author{Kun He}
\email{hekun@whu.edu.cn}
\affiliation{%
  \institution{Wuhan University}
  \city{Wuhan}
  \country{China}
}

\author{Yueming Wu}
\email{wuyueming21@gmail.com}
\affiliation{%
  \institution{Nanyang Technological University}
  \country{Singapore}
}

\author{Weisong Sun}
\email{weisong.sun@ntu.edu.sg}
\affiliation{%
  \institution{Nanyang Technological University}
  \country{Singapore}
}

\author{Ruiying Du}
\email{duraying@whu.edu.cn}
\affiliation{%
  \institution{Wuhan University}
  \city{Wuhan}
  \country{China}
}

\author{Qingchuan Zhao}
\email{cs.qczhao@cityu.edu.hk}
\affiliation{%
  \institution{City University of Hong Kong}
  \country{China}
}

\author{Yang Liu}
\email{yangliu@ntu.edu.sg}
\affiliation{%
  \institution{Nanyang Technological University}
  \country{Singapore}
}
\renewcommand{\shortauthors}{Liang et al.}



\usepackage[utf8]{inputenc}
\usepackage{textgreek}

\usepackage{amsmath,amsfonts}

\usepackage{array}
\usepackage[caption=false,font=normalsize,labelfont=sf,textfont=sf]{subfig}
\usepackage{textcomp}
\usepackage{stfloats}
\usepackage{url}
\usepackage{verbatim}
\usepackage{graphicx}

\usepackage{xcolor}
\usepackage{multirow}
\usepackage{listings}
\usepackage{booktabs}
\usepackage{mathrsfs}
\usepackage{framed}
\usepackage[ruled, boxed,linesnumbered]{algorithm2e}
\usepackage{colortbl}
\usepackage{algpseudocode}
\usepackage{amsmath}
\usepackage{enumitem}
\usepackage{graphicx}
\usepackage{float}
\usepackage{pifont}

\definecolor{light-gray}{gray}{0.80}

\definecolor{verylightgray}{rgb}{.97,.97,.97}

\lstdefinelanguage{Solidity}{
	keywords=[1]{anonymous, assembly, assert, balance, break, call, callcode, case, catch, class, constant, continue, constructor, contract, debugger, default, delegatecall, delete, do, else, emit, event, experimental, export, external, false, finally, for, function, gas, if, implements, import, in, indexed, instanceof, interface, internal, is, length, library, log0, log1, log2, log3, log4, memory, modifier, new, payable, pragma, private, protected, public, pure, push, require, return, returns, revert, selfdestruct, send, solidity, storage, struct, suicide, super, switch, then, this, throw, transfer, true, try, typeof, using, value, view, while, with, addmod, ecrecover, keccak256, mulmod, ripemd160, sha256, sha3}, 
	keywordstyle=[1]\color{blue}\bfseries,
	keywords=[2]{address, bool, byte, bytes, bytes1, bytes2, bytes3, bytes4, bytes5, bytes6, bytes7, bytes8, bytes9, bytes10, bytes11, bytes12, bytes13, bytes14, bytes15, bytes16, bytes17, bytes18, bytes19, bytes20, bytes21, bytes22, bytes23, bytes24, bytes25, bytes26, bytes27, bytes28, bytes29, bytes30, bytes31, bytes32, enum, int, int8, int16, int24, int32, int40, int48, int56, int64, int72, int80, int88, int96, int104, int112, int120, int128, int136, int144, int152, int160, int168, int176, int184, int192, int200, int208, int216, int224, int232, int240, int248, int256, mapping, string, uint, uint8, uint16, uint24, uint32, uint40, uint48, uint56, uint64, uint72, uint80, uint88, uint96, uint104, uint112, uint120, uint128, uint136, uint144, uint152, uint160, uint168, uint176, uint184, uint192, uint200, uint208, uint216, uint224, uint232, uint240, uint248, uint256, var, void, ether, finney, szabo, wei, days, hours, minutes, seconds, weeks, years},	
	keywordstyle=[2]\color{teal}\bfseries,
	keywords=[3]{block, blockhash, coinbase, difficulty, gaslimit, number, timestamp, msg, data, gas, sender, sig, value, now, tx, gasprice, origin},	
	keywordstyle=[3]\color{violet}\bfseries,
	identifierstyle=\color{black},
	sensitive=true,
	comment=[l]{//},
	morecomment=[s]{/*}{*/},
	commentstyle=\color{gray}\ttfamily,
	stringstyle=\color{red}\ttfamily,
	morestring=[b]',
	morestring=[b]"
}

\begin{document}

\title{Towards Effective Detection of Ponzi schemes on Ethereum with Contract Runtime Behavior Graph}

\begin{abstract}
  Ponzi schemes, a form of scam, have been discovered in Ethereum smart contracts in recent years, causing massive financial losses.
  Existing detection methods primarily focus on rule-based approaches and machine learning techniques that utilize static information as features.
  However, these methods have significant limitations.
  Rule-based approaches rely on pre-defined rules with limited capabilities and domain knowledge dependency. 
  Using static information like opcodes for machine learning fails to effectively characterize Ponzi contracts, resulting in poor reliability and interpretability. 
  Our research shows no significant difference between Ponzi and non-Ponzi contracts at the opcode level. 
  Moreover, relying on static information like transactions for machine learning requires a certain number of transactions to achieve detection, which limits the scalability of detection and hinders the identification of \textit{0-day} Ponzi schemes.
  
  In this paper, we propose \textsc{PonziGuard}, an efficient Ponzi scheme detection approach based on contract runtime behavior.
  Inspired by the observation that a contract's runtime behavior is more effective in disguising Ponzi contracts from the innocent contracts, \textsc{PonziGuard} establishes a comprehensive graph representation called \textit{contract runtime behavior graph} (CRBG), to accurately depict the behavior of Ponzi contracts.
  Furthermore, it formulates the detection process as a graph classification task on CRBG, enhancing its overall effectiveness.
  The experiment results show that \textsc{PonziGuard} surpasses the current state-of-the-art approaches in the ground-truth dataset, achieving a precision of 96.9\%, recall of 98.2\%, and F1-score of 97.5\%. 
  It also exhibits the highest level of interpretability among the current tools.
  We applied \textsc{PonziGuard} to Ethereum Mainnet and demonstrated its effectiveness in real-world scenarios.
  Using \textsc{PonziGuard}, we identified 805 Ponzi contracts on Ethereum Mainnet, which have resulted in an estimated economic loss of 281,700 Ether or approximately \$500 million USD.
  We also found \textit{0-day} Ponzi schemes in the recently deployed 10,000 smart contracts.

\end{abstract}

\maketitle

\renewcommand{\arraystretch}{1.3}

\section{Introduction}
\label{sec1}

With the popularity of Ethereum and the anonymity it provides, various scams have been discovered to implement themselves through smart contracts~\cite{SmartContracts}.
Ponzi schemes are one of the typical scams found in Ethereum smart contracts~\cite{DBLP:journals/fgcs/BartolettiCCS20, chainalysis}, namely Ponzi contracts, disguising as investment programs to lure users under the promise of high profits while users gain profits only if the investments made by subsequent users join the Ponzi schemes.
Ponzi schemes have been one of the biggest consumers of gas on Ethereum, heightening already bad congestion and jacking up transaction fees~\cite{cryptobriefing}.

\begin{figure}[t!]
  \centering
  \includegraphics[width=3.5in]{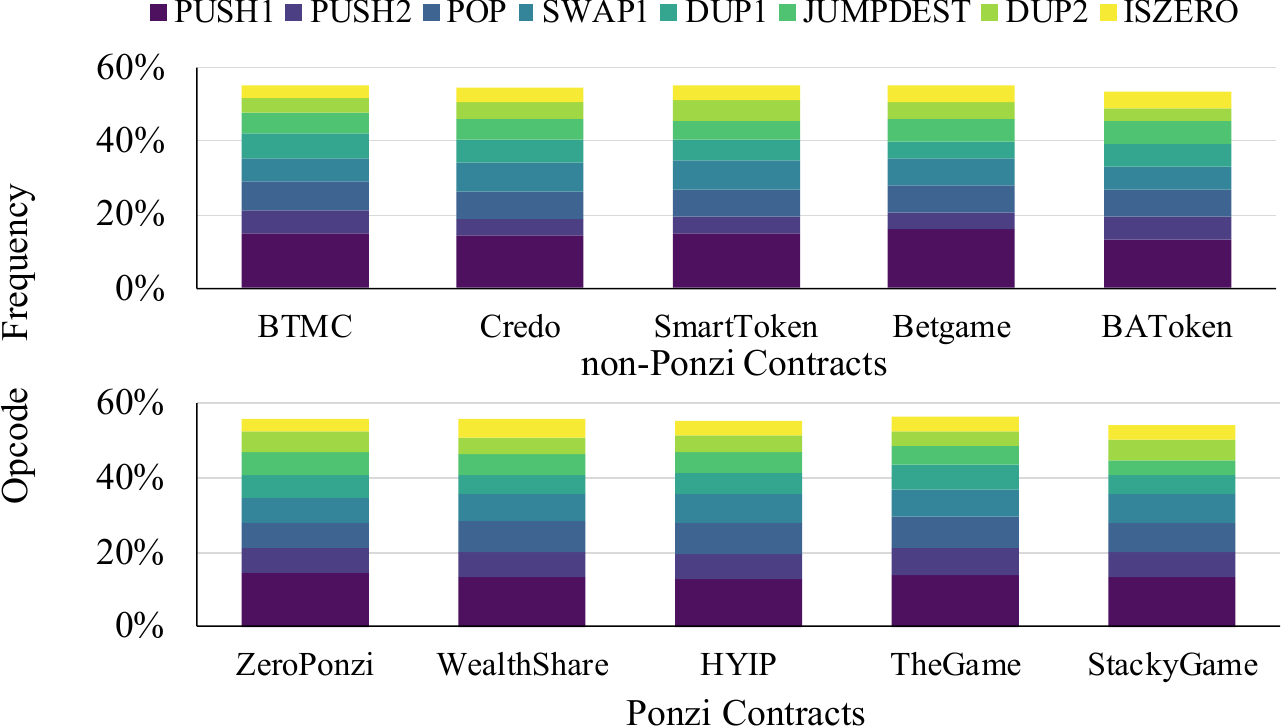}
  \caption{Opcode Frequency Distributions. The KL divergence between Ponzi and non-Ponzi contracts ranges from 0.011 to 0.018, while the KL divergence between different Ponzi contracts ranges from 0.012 to 0.016.
  }
  \label{figure 1}
\end{figure}

Several approaches have been proposed~\cite{DBLP:journals/fgcs/BartolettiCCS20, DBLP:conf/www/ChenZCNZZ18,DBLP:conf/blockchain2/JungTGG19,DBLP:conf/ijcnn/FanFXZ20,DBLP:conf/hpcc/FanXF020,DBLP:journals/ipm/FanFXC21,DBLP:conf/IEEEscc/LouZC20,DBLP:conf/qrs/SunXY020,DBLP:conf/sigmetrics/ChenLSHWWL21,DBLP:conf/blocksys/YuJX0X21} to detect Ponzi contracts on Ethereum.
Rule-based approaches~\cite{DBLP:journals/fgcs/BartolettiCCS20,DBLP:conf/sigmetrics/ChenLSHWWL21,DBLP:conf/qrs/SunXY020} require domain knowledge on Ponzi schemes and can hardly cover all possible scenarios based on the existing known Ponzi contracts, which limits their capability to detect Ponzi contracts that fall outside the scope of the rules.
Other detection approaches use static information such as opcode frequency and transactions for machine learning models to improve detection capabilities\cite{DBLP:conf/www/ChenZCNZZ18,DBLP:conf/blocksys/YuJX0X21,DBLP:conf/blockchain2/JungTGG19,DBLP:conf/IEEEscc/LouZC20, DBLP:conf/ijcnn/FanFXZ20,DBLP:conf/hpcc/FanXF020,DBLP:journals/ipm/FanFXC21,10448439,10.1145/3571847}.
However, this static information has a low correlation with Ponzi schemes themselves, and these approaches fail to effectively characterize the Ponzi contracts, resulting in poor reliability and interpretability.
For instance, Figure~\ref{figure 1} shows the frequency distributions of some most frequently used operations in some Ponzi contracts and non-Ponzi contracts.
These operations are predominantly stack operations and do not capture the characteristics of Ponzi contracts.
The \textit{Kullback-Leibler Divergence} (KL divergence) calculated from Figure~\ref{figure 1} measures the difference between two frequency distributions.
It can be concluded from the KL divergence that the distributions of opcode frequency exhibit low differences between Ponzi and non-Ponzi contracts, and no substantial similarities between different Ponzi contracts by comparison.
Moreover, those approaches utilizing Ethereum transactions cannot detect \textit{0-day} Ponzi contracts, i.e., having none real transactions.

To address this gap, we delved deeper into the behaviors of Ponzi contracts at runtime and found that the contract runtime information provides more valuable insights into the unique characteristics of Ponzi contracts.
We will discuss this insight in more detail in Section~\ref{section Background And Insight}.
Motivated by this observation, we propose a comprehensive graph representation called \textit{contract runtime behavior graph} (\textbf{CRBG}) to characterize the runtime behaviors of Ponzi contracts.

In this paper, we propose \textsc{PonziGuard}, an effective Ponzi scheme detection approach based on CRBG.
Specifically, we perform static analysis on the smart contract and leverage the acquired insights to generate transaction sequences that mimic typical investment behavior in a Ponzi scheme.
We invoke the smart contract with these transaction sequences and conduct dynamic taint analysis during the contract's execution to gather runtime information.
Then, we construct CRBG based on the contract runtime information and empower Graph Neural Networks (GNNs) for CRBG analysis.
We formulate the detection of Ponzi contracts as a graph classification task.
We have experimentally validated the effectiveness of CRBG and conducted comparative experiments on a ground-truth dataset to evaluate the performance of \textsc{PonziGuard}.
We further applied \textsc{PonziGuard} to Ethereum Mainnet to evaluate the effectiveness of our approach in real-world scenarios.
The dataset and experimental results are publicly available online\footnote{\url{https://github.com/PonziDetection/PonziGuard}}.
In summary, this paper makes the following contributions.

\begin{itemize}[leftmargin=4mm, itemindent=0mm]
  \item We propose \textsc{PonziGuard}, an efficient approach for detecting Ponzi schemes on Ethereum.
  It does not require any domain knowledge and on-chain transaction.
  It can identify \textit{0-day} Ponzi schemes before any economic losses occur.
  \item We introduce CRBG, a comprehensive graph representation for effectively characterizing the behaviors of Ponzi contracts.
  We model the detection of Ponzi contracts as a graph classification task and prove that CRBG is effective in disguising the Ponzi contracts from the innocent contracts.
  \item We propose a strategy for generating contract invoke sequences based on function properties and data dependencies, enabling us to mimic typical investment behavior in a Ponzi scheme.
  We also design a dynamic taint engine to collect contract runtime behavior, which is essential for constructing CRBG.

  \item Experimental results show that \textsc{PonziGuard} outperforms the current state-of-the-art approaches on the ground-truth dataset and is also effective in real-world scenarios.
  We found 805 Ponzi contracts using \textsc{PonziGuard} out of 14,000,000 Ethereum Mainnet blocks which have resulted in an estimated economic loss of 281,700 Ether or approximately \$500 million USD.
  We also found \textit{0-day} Ponzi schemes in the recently deployed 10,000 smart contracts.
\end{itemize}

This paper is an extended version of our previous work~\cite{PonziGuard} published in IEEE/ACM International Conference on Software Engineering (ICSE) 2024.
We significantly enhanced the previous conference version in the following aspects:
i) we added a static code analysis framework prior to generating transactions for collection of information such as data dependencies and function properties, to guide the generation of transaction sequences (\S\ref{section Static Code Analysis});
ii) we updated our transaction sequence generation algorithm, using information from static analysis and some heuristics to generate function invoke chains that mimic investment behavior of Ponzi schemes (\S\ref{section Transaction Sequence Generation});
iii) we performed additional processing on CRBG including graph pruning and subgraph joining to achieve improved experimental results (\S\ref{section CRBG Construction});
iv) we conducted the comparison experiment using a larger and more recent dataset, and added two SOTA works for comparison (\S\ref{section Effectiveness of PonziGuard});
v) we conducted an interpretability experiment to visually explain how our CRBG works in the detection of Ponzi schemes (\S\ref{section Interpretability});
vi) we conducted an ablation experiment to illustrate the impact of the graph pruning and joining processing that we integrated into CRBG (\S\ref{section Effectiveness of graph pruning and joining});
vii) we explained the reasons behind the short lifetime of the 805 Ponzi contracts we identified, including how they were terminated (\S\ref{section Performance in real-world scenarios});
viii) we conducted an experiment to detect newly deployed \textit{0-day} Ponzi schemes on Ethereum, showcasing our tool's ability to detect \textit{0-day} Ponzi schemes (\S\ref{section Detecting 0-day Ponzi Schemes}).

\section{Background And Insight}
\label{section Background And Insight}
In this section, we introduce some necessary backgrounds and discuss our insight into utilizing the contract runtime behavior graph (CRBG) to detect Ponzi schemes.

\subsection{Ethereum Smart Contracts}
Ethereum smart contracts are programs running on top of Ethereum.
They can be written in several programming languages, including Solidity, Viper, and Serpent.
To deploy smart contracts on the blockchain, they need to be compiled into bytecode and then submitted to the blockchain with transactions.
Once deployed on-chain, the contracts become immutable and the implementation of their logic relies on message calls from transactions.
When invoked by a transaction, contracts will be executed in Ethereum Virtual Machine (EVM), a stack-based architecture~\cite{Ethereum}.
There are three areas to store data in EVM:
\begin{itemize}[leftmargin=4mm, itemindent=0mm]
  \item \textbf{Stack}: The stack is an object for basic stack operations in EVM.
        Data is pushed or popped from the top of the stack through instructions.
  \item \textbf{Memory}: The memory is a simple word-addressed byte array. 
  It is used for temporary data storage, transfer of arguments and return values, and code copying~\cite{Ethereum}.
  The data in the memory comes from the stack or the external environments.
  \item \textbf{Storage}: Unlike the memory and stack that are volatile, the storage is non-volatile and maintained as part of the smart contract state.
        Variables in the storage region are called state variables, and they are persistent variables stored in the form of key-value pairs.
        Transactions can update the state variables of smart contracts by invoking the execution of contracts in EVM.
\end{itemize}

\subsection{Ponzi Schemes}
\label{section Ponzi Schemes}
A Ponzi scheme is an investment fraud that involves the payment of purported returns to existing investors from funds contributed by new investors~\cite{SEC}.
It is a classic fraud that originated at least 150 years ago and now appears on blockchains~\cite{DBLP:journals/fgcs/BartolettiCCS20}.
Leveraging smart contracts, Ponzi schemes become more threatening and stealthy than ever and have grabbed a huge amount of profits on the blockchain~\cite{chainalysis}.

\begin{lstlisting}[language = Solidity, escapechar=!, caption = Motivating Example, label = lst 1, float=t]
function enter(){
  if(msg.value <1/100 ether){
    msg.sender.send(msg.value);
    return;}
  uint amount = msg.value;
  uint idx = persons.length;
  persons.length += 1;
  persons[idx].etherAddress = msg.sender;
  persons[idx].amount = amount;}

function pay(){
  while(this.balance > persons[payoutIdx].amount / 100 * 500){
    uint transactionAmount = persons[payoutIdx].amount / 100 * 500;
    persons[payoutIdx].etherAddress.send(transactionAmount);
    payoutIdx += 1;}}
\end{lstlisting}

\begin{figure}[t!]
  \centering
  \includegraphics[width=5.7in]{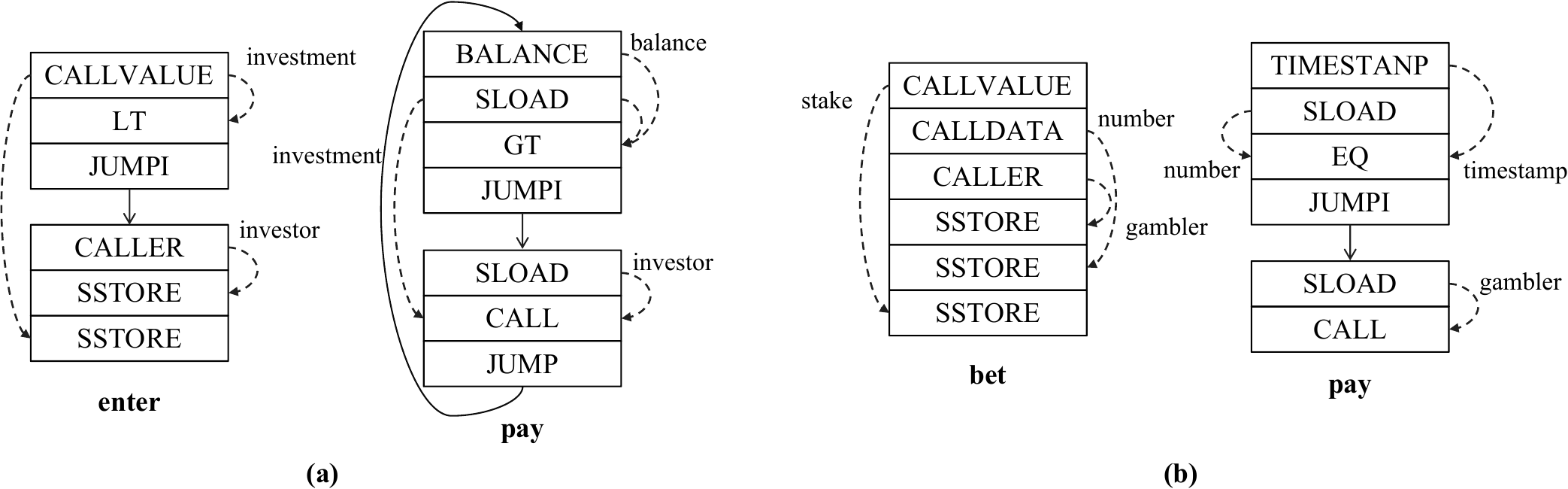}
  \caption{Contract Runtime Behaviors. (a) is obtained from the Ponzi contract in Listing~\ref{lst 1}, and (b) is obtained from a gambling contract (non-Ponzi contract).}
  \label{figure 2}
\end{figure}

\subsubsection{Code Example}
Listing~\ref{lst 1} shows a code snippet of a typical Ponzi contract.
The snippet comprises two functions, namely \texttt{enter()} and \texttt{pay()}, where \texttt{enter()} is responsible for receiving Ether from investors and \texttt{pay()} handles the redistribution of Ether.
This contract promises investors very high return rates (Line 13) in exchange for their initial investment.
The promised returns are paid out of new investments to attract additional investors until the scammers close up their scam and abscond with the illicit profits.
Without legitimate earnings, a Ponzi scheme needs a steady stream of new investors to keep it running, otherwise, it will inevitably collapse and let the vast majority of participants bear the loss~\cite{ARTZROUNI2009190}.

\subsubsection{Criteria}
Based on some previous studies~\cite{DBLP:journals/fgcs/BartolettiCCS20,SEC,DBLP:conf/sigmetrics/ChenLSHWWL21} and our analysis of known Ponzi contracts, we have developed explicit criteria for objectively identifying Ponzi contracts in our study.
Our proposed criteria include:
\begin{itemize}[leftmargin=4mm, itemindent=0mm]
  \item A Ponzi contract must incorporate at least two explicit behavioral logics: investment and reward.
  This criterion excludes contracts that receive cryptocurrency but provide users with assets through external markets such as real-world trades or auctions that utilize cryptocurrency for payments.
  \item The assets of a Ponzi contract must come from a multitude of investors rather than a specific source. 
  This means that Ponzi contracts have no sources of income other than attracting investments.
  This criterion excludes contracts specifically designed to fulfill certain functions, such as enterprises distributing incentives to employees.
  \item In a Ponzi contract, all the investors are promised rewards that are typically expected to exceed their initial investment, although the implementation of these rewards is contingent upon attracting further investments.
  In other words, as long as there are constant new investments, everyone can theoretically reap the rewards.
  This criterion excludes the contracts that are likely to be mistaken for Ponzi contracts, such as gambling and puzzle contracts.
  In such contracts, not all users are promised rewards as they would be in a Ponzi contract. (There are always losers in gambling or puzzle games.)
\end{itemize}

\subsection{Contract Behaviors and Our Insight}
Through our proposed criteria, we can observe that the most crucial distinction between Ponzi contracts and benign contracts lies in their behavioral characteristics, such as the investment and reward logics and the flow of Ether, rather than specific transaction or instruction-level statistics.
Therefore, in this section, we explore the contract runtime behaviors, trying to find an effective representation of these behaviors.

We invoke the smart contracts, gather their runtime information, and construct graphs based on this information, as depicted in Figure~\ref{figure 2}.
It is important to note that the graphs in Figure~\ref{figure 2} have been intentionally simplified to highlight the core logic of the contracts for the sake of clarity.
The left graph in Figure~\ref{figure 2}(a) depicts the investment behavior of the \texttt{enter()} function within the Ponzi contract shown in Listing~\ref{lst 1}. 
This contract first utilizes the \texttt{CALLVALUE} and \texttt{LT} operations to compare the Ether amount provided by investors (corresponding to Line 2 in Listing~\ref{lst 1}), and then utilizes \texttt{SSTORE} to store the investment amount and the address of the investors (Line 8 and Line 9).
As it only relies on the comparison of the investment amount as the condition for receiving Ether, the source of Ether for the contract is not restricted to a specific address but encompasses all investors, which aligns with our second criterion for Ponzi contracts.
The right graph in Figure~\ref{figure 2}(a) depicts the reward behavior of the \texttt{pay()} function within the Ponzi contract shown in Listing~\ref{lst 1}.
It first uses \texttt{SLOAD} to load the investment amount of the investor and calculates the promised reward (corresponding to Line 12 in Listing~\ref{lst 1}).
If the contract balance is deemed sufficient to cover the reward, as determined through the comparison using \texttt{BALANCE} and \texttt{GT}, it proceeds to load the investor's address and completes the transfer (Line 14).
Since this reward process iterates in a loop where the only condition for transferring Ether to investors is a sufficient contract balance, it can be inferred that every investor can potentially receive a reward as long as there is a continuous influx of investors, which aligns with our third criterion.
The behaviors presented by these two graphs of Figure~\ref{figure 2}(a) also satisfy our first criterion for Ponzi contracts.
For comparison, consider Figure~\ref{figure 2}(b), which represents the bet and pay behaviors of a gambling contract\footnote{0x4f9048d95616dbf7acc16fc4179f5ac6ee37bce6}. 
In this case, the contract only rewards the gambler whose pre-selected number precisely matches the current timestamp. 
While this gambling contract fulfills the first and second criteria, it falls short of meeting our third criterion for Ponzi contracts.

In conclusion, these graphs depict the behaviors of the Ponzi contract as reflected in its source code and fulfill the criteria we have proposed, distinguishing it from benign contracts.
This demonstrates that the graphs we constructed have the capability to effectively reveal the distinctive behavioral traits of Ponzi contracts.
We refer to these graphs as contract runtime behavior graphs (CRBG).
The illustrated graphs in Figure~\ref{figure 2} serve as a preliminary illustration for clarity, while a more comprehensive description of CRBG can be found in Section~\ref{section CRBG Construction}.

\begin{figure*}[t!]
  \centering
  \includegraphics[width=5.3in]{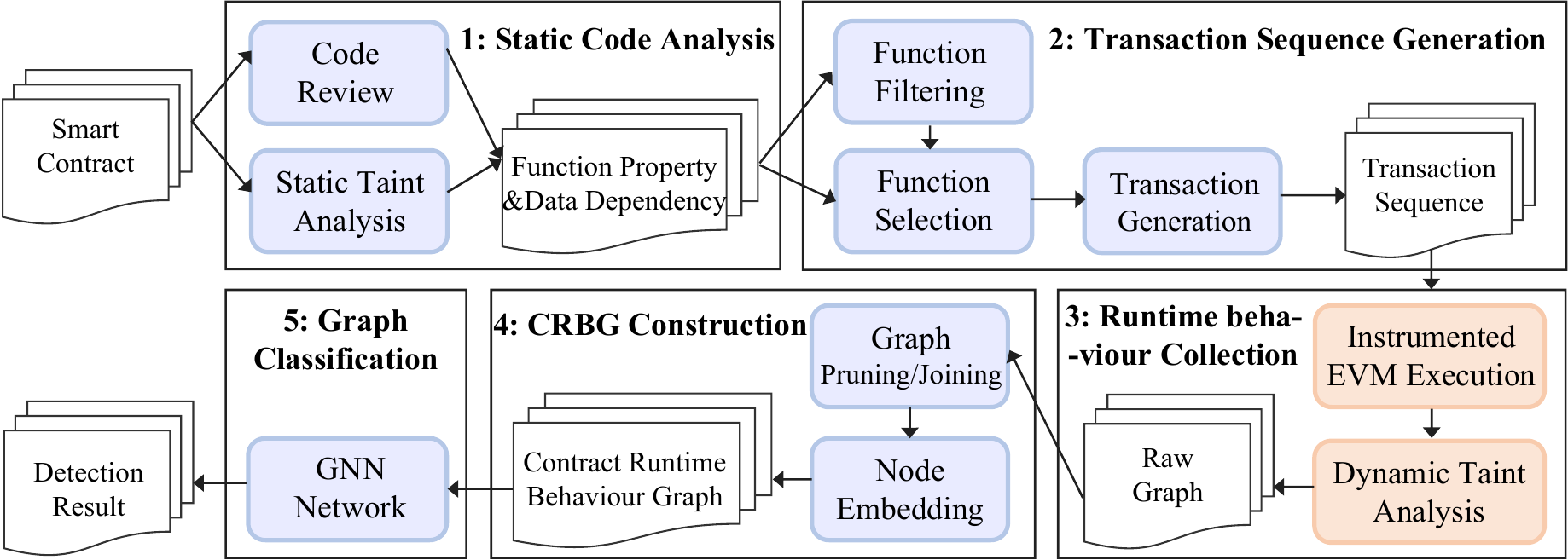}
  \caption{Overview of \textsc{PonziGuard}.}
  \label{figure 3}
\end{figure*}

\section{PonziGuard}
We first give an overview of \textsc{PonziGuard}.
Then, we describe each step in detail.

\subsection{Overview}

As shown in Figure~\ref{figure 3}, we conduct code review and static analysis on the smart contract to extract function properties and identify data dependencies.
Based on this gathered information, we conduct function filtering and selection to generate function invocation chains that mimic the investment behavior seen in a Ponzi scheme.
Subsequently, we generate the corresponding transaction sequences to invoke the contract.
Leveraging the dynamic taint engine, we collect runtime information during the contract execution within the instrumented EVM, creating raw graphs that encompass runtime control flow and data flow.
After applying graph pruning and subgraph joining, we obtain the contract runtime behavior graph (CRBG).
The CRBG serves as input for training a Graph Neural Network (GNN) designed to classify the graph. 
Consequently, we transform the Ponzi scheme detection into a graph classification task.

\subsection{Static Code Analysis}
\label{section Static Code Analysis}

In order to enhance the likelihood of triggering the behavior of Ponzi contract, we conduct static analysis to provide valuable information for generating transaction sequences.
Specifically, we first examine the contract's source code to extract essential function properties, including the function name, visibility, and whether they are declared payable (i.e., capable of receiving Ether).
Then, we leverage \textsc{slither}~\cite{feist2019slither}, a static analysis framework designed for smart contracts, to extract read and write operations of each function on state variables.
Subsequently, we organize the data dependencies related to the functions within the contract.
For instance, if function \texttt{A} writes to state variable $\alpha$ and function \texttt{B} reads from state variable $\beta$, where $\beta$ is dependent on or exactly $\alpha$, we include function \texttt{B} to the list of functions with a data dependency relationship on function \texttt{A}.

\begin{algorithm}[]
  \SetAlgoLined
  \SetAlgoNlRelativeSize{-1}
  \SetKwInOut{Input}{Input}
  \SetKwInOut{Output}{Output}
  \caption{Transaction Sequences Generation.}
  \label{algorithm 1}
  \Input{$F_{kws}, F_{payable}, F_{w}, F_{all}, Max$}
  \Output{$TxSequences$}
  \SetKwProg{try}{try}{:}{}
  \SetKwProg{catch}{catch}{:}{end}
  $g$ $\gets$ 0\\
  $TxSequences \gets init()$\\
  \While{$g$ \textless $Max$}{ 
    $txs$ $\gets$ $init()$\\
    $func$ $\gets$ $randomChoose$($F_{kws}$, $F_{payable}$, $F_{w}$)\\
    \If{$func$ is empty}{
      $func$ $\gets$ $randomChoose$($F_{all}$)
    }
    $tx \gets generateTransaction(func)$\\
    $txs.add(tx)$\\
      \While{len($txs$) \textless  len($F_{all}$)}{
        $F_{dep}$ $\gets$ \textit{$getDependency$}($txs$)\\
        \If{$F_{dep}$ is not empty}{
            $func$ $\gets$ $randomChoose$($F_{dep}$, $F_{all}$)
        }
        \Else{$func$ $\gets$ $randomChoose$($F_{all}$)}
    
        $tx \gets generateTransaction(func)$\\
        $txs.add(tx)$}
    $TxSequences.add(txs)$\\
    $g \gets g + 1$
  }
\end{algorithm}

\subsection{Transaction Sequence Generation}
\label{section Transaction Sequence Generation}

We utilize the extracted function properties and data dependencies to generate transactions, thereby simulating the investment behavior of the Ponzi contract and triggering its functionality.
Due to the presence of state variables in smart contracts, prior transactions can influence the execution of subsequent ones. 
Therefore, achieving successful investment may necessitates multiple transactions. 
Accordingly, we generate transaction sequences to interact with the contract.
The process of generating transaction sequences is outlined in Algorithm~\ref{algorithm 1}. 
Initially, we filter out functions that are irrelevant to the contract's (Ponzi) behavior, such as view functions and pure functions, and categorize specific functions into distinct groups
For instance, in Algorithm~\ref{algorithm 1}, $F_{kws}$ denotes functions whose names contain keywords (such as \texttt{invest, enter, init, deposit}); $F_{payable}$ encompasses payable functions, including the fallback function; $F_{w}$ encompasses functions capable of altering state variables, while $F_{all}$ encompasses all functions in the contract.
We use heuristics to select the initial function to invoke (Lines 4-10).
Specifically, we give priority to payable functions because the investment function is typically declared as payable.
Additionally, we prioritize functions whose names contain specific keywords because they are more likely to encapsulate the logic of Ponzi contract investment.
Once the first transaction is generated (Line 10), we proceed to complete the transaction sequences in a Read-After-Write order (Lines 11-22), which is a common practice to create meaningful function invocation chains.
The function $randomChoose()$ selects a function randomly from its arguments (Line 5), while $generateTransaction()$ is responsible for creating a valid transaction based on the chosen function (Line 9). 
To achieve this, $generateTransaction()$ initially analyzes the contract's ABI (Application Binary Interface), then randomly selects values within the valid input range for fixed data types, such as uint256. 
For non-fixed data types like string, it determines a positive number as the data length and generates an input of that length. 
Additionally, for payable functions, $generateTransaction()$ employs a continuously increasing flow of Ether attached to transactions to facilitate the activation of specific behaviors of Ponzi contracts, such as investment and reward.
The function $getDependency()$ retrieves functions that have data dependencies on the provided transactions (Line 12).

\subsection{Runtime Behavior Collection}
\label{section Runtime Behavior Collection}

We use the generated transaction sequences to trigger contract execution, and collect the contract runtime behavior information.
To achieve this, we design a taint engine for EVM to perform dynamic taint analysis and gather runtime details of smart contracts.
Dynamic taint analysis, a widely-used program analysis technique~\cite{DBLP:conf/sp/SchwartzAB10}, utilizes predefined taints to track program execution and observe runtime data flow and control flow.

\subsubsection{Taint Sources and Sinks}
\label{section Taint Sources and Sinks}

As shown in Table~\ref{table 1}, we have selected some operations as taint sources to introduce taint data.
These operations will push some external data into the stack or memory, such as \texttt{CALLER}, \texttt{CALLVALUE}, \texttt{CALLDATALOAD}, \texttt{CALLDATACOPY}, which are related to the transaction sender and arguments, and \texttt{TIMESTAMP}, \texttt{BLOCKHASH}, which are related to the blockchain environment.
In addition, we also consider some operations related to the contract itself, such as \texttt{BALANCE} and \texttt{ADDRESS}.
Data derived from these sources is marked as tainted, while other data is marked as untainted. 
Regarding taint sinks, we select some meaningful operations as the location to check the flow of taint data. 
These operations either take taint data as their arguments (e.g., \texttt{GT}, \texttt{CALL} and \texttt{SSTORE}), or load taint data and push it into the stack (e.g., \texttt{MLOAD} and \texttt{SLOAD}).

\subsubsection{Taint Propagation}
\label{section Taint Propagation}

To achieve the taint propagation, we implement the taint engine that encompasses the components such as a taint stack, a taint memory, and a taint storage.
Each slot of the taint stack contains a taint that marks the corresponding slot in the EVM stack.
Since the EVM memory is a byte array, each taint in the taint memory is responsible for a byte in the EVM memory.
Both the taint stack and taint memory are volatile regions that are freed and allocated at the start of each new transaction. 
In contrast, the EVM storage is non-volatile and stores state variables in key-value pairs. 
In these key-value pairs, a 32-byte address calculated from the state variable is stored as the key, and the state variable is stored as the value.
We maintain the taint storage in the same structure, with the address of the state variable as the key and the taint as the value.
As storage is non-volatile, the taint storage is kept until all transactions are completed, as part of the Ethereum world state.
In general, when one operand of an arithmetic operation is tainted, the result of the operation is also tainted regardless of the other operands.
The implementation of the taint engine enables us to capture and trace the data flow throughout the contract execution.

\subsubsection{Raw Graphs}
\label{section Raw Graphs}

We gather the information obtained in the contract execution and construct a raw graph that integrates the control flow and data flow of the contract runtime.
The nodes of the graph are the operations executed during runtime, and we add control flow and data flow edges as the graph edges.
The control flow edges are categorized into six types, with the most common type being the adjacent edge.
This edge connects two operations whose program counters differ by only 1, indicating that they are executed in a successive manner.
The other types of control flow edges include the jump edge, which connects \texttt{JUMP}(\texttt{I}) and the operation executed after the jump, as well as the call, return, and creation edges that similarly connect the corresponding operation (e.g., \texttt{CALL}, \texttt{RETURN}, \texttt{CREATE}) and its successor.
Regarding the data flow edges, we follow the principle of adding edges from taint sources to taint sinks, representing the propagation of taint data. 
There are eight kinds of data flow edges according to the taint sources in Table~\ref{table 1}.

\begin{figure}[t!]
  \centering
  \includegraphics[width=4.5in]{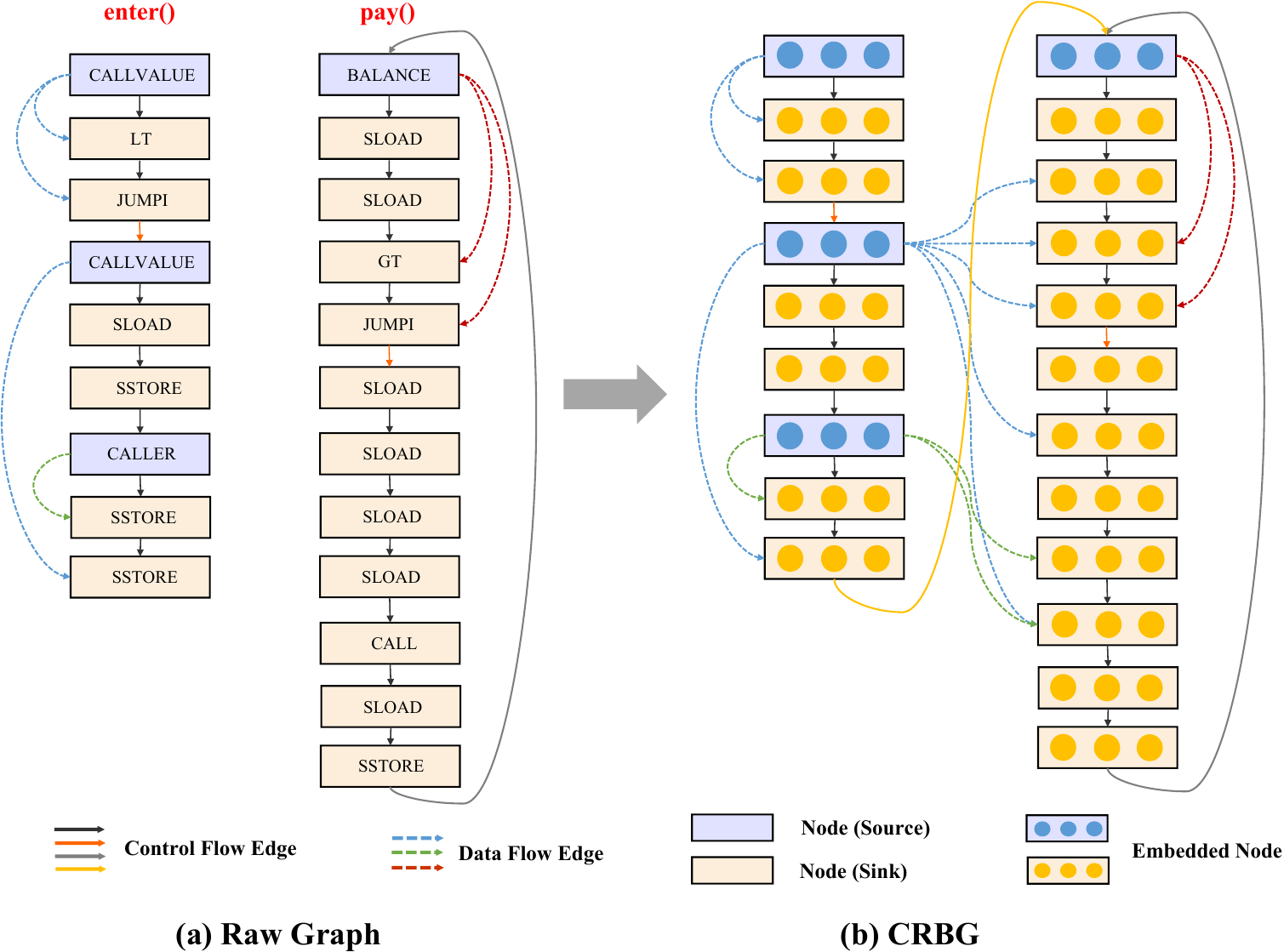}
  \caption{CRBG Construction. (a) shows the raw graphs from dynamic taint analysis, each graph corresponds to a invocation. (b) is the CRBG with better node embeddings and more comprehensive runtime information. We simplify the graphs by keeping only the nodes representing the taint sources and sinks that capture the main logic of the functions for the convenience of display.}
  \label{figure 4}
\end{figure}

Figure~\ref{figure 4}(a) shows examples of the output graphs obtained from the dynamic taint analysis.
The two graphs presented in Figure~\ref{figure 4}(a) are the result of invoking \texttt{enter()} and \texttt{pay()} within the contract shown in Listing~\ref{lst 1}.
For the sake of clarity, only the nodes representing the taint sources and sinks that capture the main logic of the functions are included in these simplified graphs shown in Figure~\ref{figure 4}(a).

\renewcommand{\arraystretch}{1} 

\begin{table}[t!]
  \caption{Taint Sources and Sinks.}
  \label{table 1}
  \centering
  \begin{tabular}{l|c}
  \toprule
  \multicolumn{1}{c|}{\textbf{Sources}}                                                                        & \textbf{Opcode Type}   \\ \midrule
  \rowcolor[HTML]{EFEFEF} 
  \begin{tabular}[c]{@{}l@{}}\scalebox{0.95}{CALLVALUE/CALLDATASIZE/CALLER/ORIGIN/CALLDATALOAD/CALLDATACOPY}\end{tabular} & Transaction Related    \\
  TIMESTAMP / BLOCKHASH                                                                                          & Blockchain Environment \\
  \rowcolor[HTML]{EFEFEF} 
  BALANCE / ADDRESS                                                                                              & Contract Related       \\ \midrule
  \multicolumn{1}{c|}{\textbf{Sinks}}                                                                          & \textbf{Opcode Type}   \\ \midrule
  \rowcolor[HTML]{EFEFEF} 
  EQ / LT / SLT / GT / SGT                                                                                             & Comparison             \\
  MSTORE / MSTORE8 / MLOAD                                                                                         & Memory Related         \\
  \rowcolor[HTML]{EFEFEF} 
  SSTORE / SLOAD                                                                                                 & Storage Related        \\
  \begin{tabular}[c]{@{}l@{}}CALL / CALLCODE / DELEGATECALL / STATICCALL\end{tabular}                             & Call                   \\
  \rowcolor[HTML]{EFEFEF} 
  JUMPI                                                                                                        & Jump                   \\ \bottomrule
  \end{tabular}
\end{table}

\renewcommand{\arraystretch}{1.3}

\subsection{CRBG Construction}
\label{section CRBG Construction}

\begin{figure}[t!]
  \centering
  \includegraphics[width=4.3in]{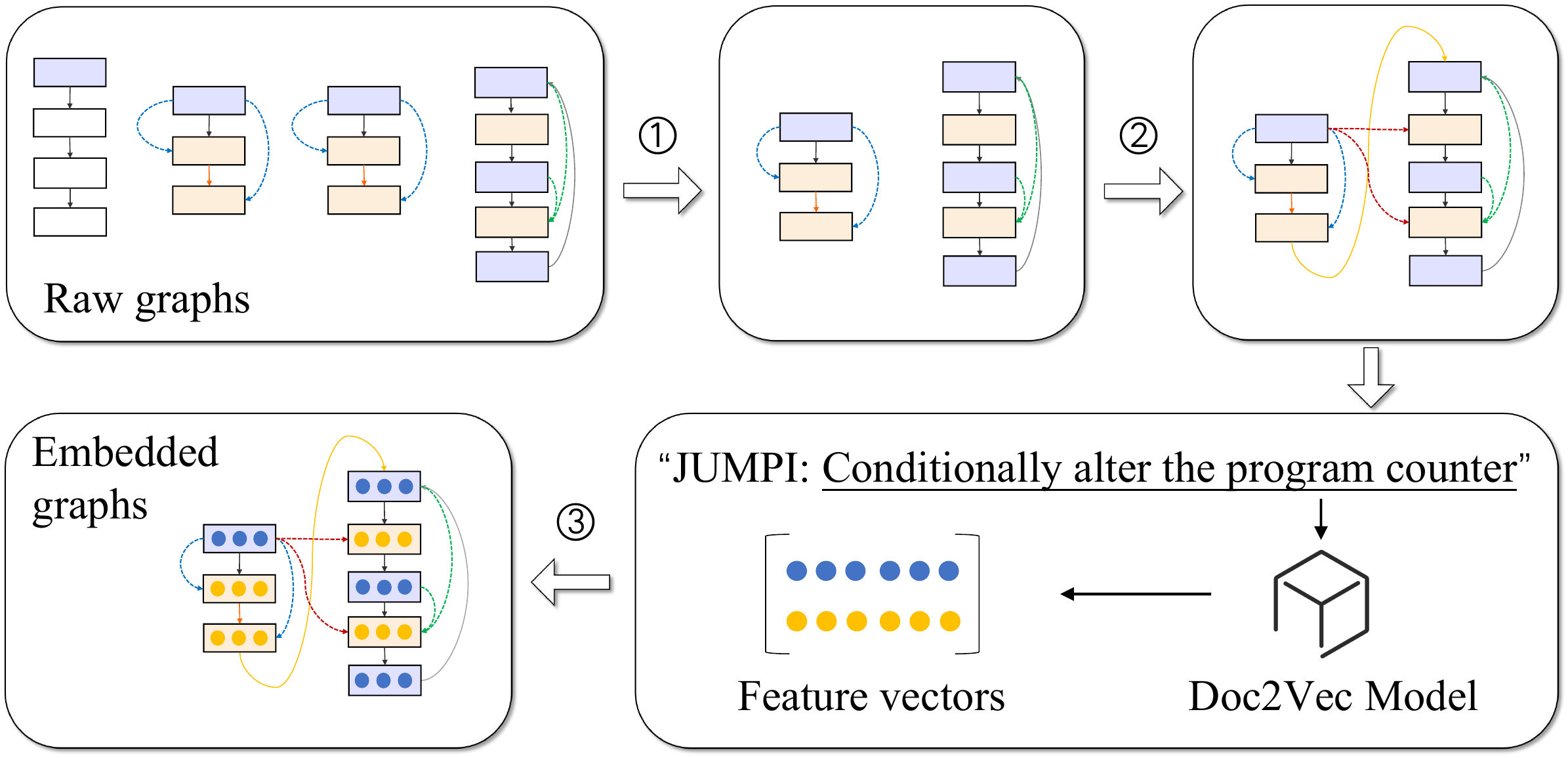}
  \caption{Processing raw graphs. \ding{172} denotes the pruning of raw graphs. \ding{173} denotes the joining of independent subgraphs. \ding{174} denotes the node feature embeddings.}
  \label{figure 17}
\end{figure}

In this section, we illustrate why the raw graph obtained from the runtime behavior collection stage is not suitable for training an effective model and how we process these raw graphs to improve their representation of the behavioral characteristics of the Ponzi contract, making them suitable inputs for the graph neural network.
The entire process is outlined in Figure~\ref{figure 17}.

\subsubsection{Mitigating Graph Data Redundancy}

As outlined in Section~\ref{section Runtime Behavior Collection}, we generate transactions to invoke the contract and execute its designated behavior patterns. 
Each transaction triggers the contract to execute once, generating a corresponding graph structure.
To maximize the activation of Ponzi scheme behavior patterns, we generate multiple transaction sequences for each contract. 
Consequently, numerous redundant graphs occur, corresponding to repeated contract executions, failed executions due to contract assertions or invalid inputs, and executions unrelated to Ponzi scheme behavior patterns.

To mitigate graph data redundancy, we adopted two strategies to prune the raw graphs. 
The first strategy involves pruning based on contract behavior. 
This entails removing graphs from failed executions and executions that are not related to Ponzi scheme behavior patterns.
Specifically, if an execution does not contain the opcode \texttt{CALLVALUE} and \texttt{CALLER}, it is unlikely to represent an investment operation of a Ponzi scheme. 
Similarly, if the execution lacks comparison (such as \texttt{LT}, \texttt{GT}, \texttt{EQ}) on the state variables, it is unlikely to be a reward operation of a Ponzi scheme. 
Therefore, we remove from the graphs any executions that are neither investment nor reward behaviors. 
Additionally, we remove graphs without the \texttt{SSTORE} opcode, as it is one of the most commonly used opcodes to modify the contract state, and executions without \texttt{SSTORE} often indicate that they ended in failure.
The second strategy involves pruning similar graphs. 
We compute the cosine similarity of node and edge features between each pair of graphs to assess their similarity. 
Graphs surpassing a specified threshold of similarity are deemed similar.
Among similar graphs, we retain only one and discard the others.

\subsubsection{Joining Independent Graphs and Enhancing Data Flow Integrity}

After the invocation, a contract may correspond to multiple graphs, as each transaction can invoke the contract and generate a graph, as depicted in Figure~\ref{figure 17}.
However, an individual graph may not be sufficient to fully capture the behavior of the contract.
For instance, in Figure~\ref{figure 4}(a), each graph only depicts a single stage of the contract (i.e., the investment stage for \texttt{enter()} and the reward stage for \texttt{pay()}), and neither of these graphs alone can conclusively determine that it is a Ponzi contract. 
Moreover, the data flow of the contract is isolated among graphs.
Since smart contracts have persistent variables, there may also be data flow across transactions, which cannot be captured by individual graphs.

To address these issues, we connect all graphs of the same contract sequentially using a new type of edge called connection edge. 
This creates a connected graph that can better represent the behavior of the contract across multiple transactions.
Furthermore, we complete the \textit{across-transaction} data flow among previously independent graphs using taint storage.
The taint storage records the taint status of variables at the end of each transaction, and this information is used to propagate taints to subsequent transactions. 
With these enhancements, we are able to capture data flow that spans multiple transactions and more accurately analyze the behavior of smart contracts.

\subsubsection{Enriching Node Features}
In the raw graph, nodes are distinguished by the type of operation they include. 
This results in each node being represented by a 139-dimensional one-hot vector (corresponding to 139 unique operations), with a single non-zero entry corresponding to the type of operation. 
However, one-hot vectors do not capture any information about the relationships between nodes in the graph, which are crucial for understanding its structure and properties. 
To improve the classification accuracy, we need better node embeddings that can capture these relationships.

We noticed that there is an introduction for each operation in the Ethereum Yellow Paper~\cite{Ethereum} as exemplified in Table~\ref{table 2}.
In Table~\ref{table 2}, $\alpha$ represents the additional items placed on the stack, while $\delta$ represents the items removed from the stack~\cite{Ethereum}. 
The description section explains how the operation works in text, and shows how it operates the data in the EVM in the formula.
To embed the nodes, we first remove the formula in the description section and keep only the text explanation to preserve the functional information of the operation. 
As shown in Figure~\ref{figure 17}, for each node, we use Doc2Vec~\cite{le2014distributed}, a model for generating embeddings of variable-length pieces of text, to convert the text explanation into a 100-dimensional vector, which we stitch together with $\alpha$ and $\delta$ to form the node feature. 
After that, we have completed the construction of CRBG which will be labeled for model training later.
Figure~\ref{figure 4}(b) shows the constructed CRBG after the raw graphs in Figure~\ref{figure 4}(a) were preprocessed. 
The CRBG in Figure~\ref{figure 4}(b) has better node embeddings, more comprehensive contract runtime information, and can better characterize the contracts.

\renewcommand{\arraystretch}{1} 

\begin{table}[t!]
  \caption{Introduction of JUMPI.}
   \label{table 2}
   \centering
    \begin{tabular}{@{}c|c|c|c|c@{}}
    \toprule
    \textbf{Value} & \textbf{Mnemonic} & \textbf{$\delta$} & \textbf{$\alpha$} & \textbf{Description}                                                                                      \\ \midrule
    0x57  & JUMPI    & 2 & 0 & \begin{tabular}[c]{@{}c@{}}Conditionally alter the program counter.\\ \\ $\mathrm{J_{JUMPI}(\mu)\equiv 
        \left\{
        \begin{aligned}
        \mu_s[0]\quad if\; \mu_s[1]\neq0\\
        \mu_{pc}+1\quad otherwise\nonumber
        \end{aligned}
        \right.}$\end{tabular}\\ \bottomrule
    \end{tabular}

\end{table}

\renewcommand{\arraystretch}{1.3}

\subsection{Graph Classification}
\label{section Graph Classification}

\begin{figure}[t!]
  \centering
  \includegraphics[width=3.5in]{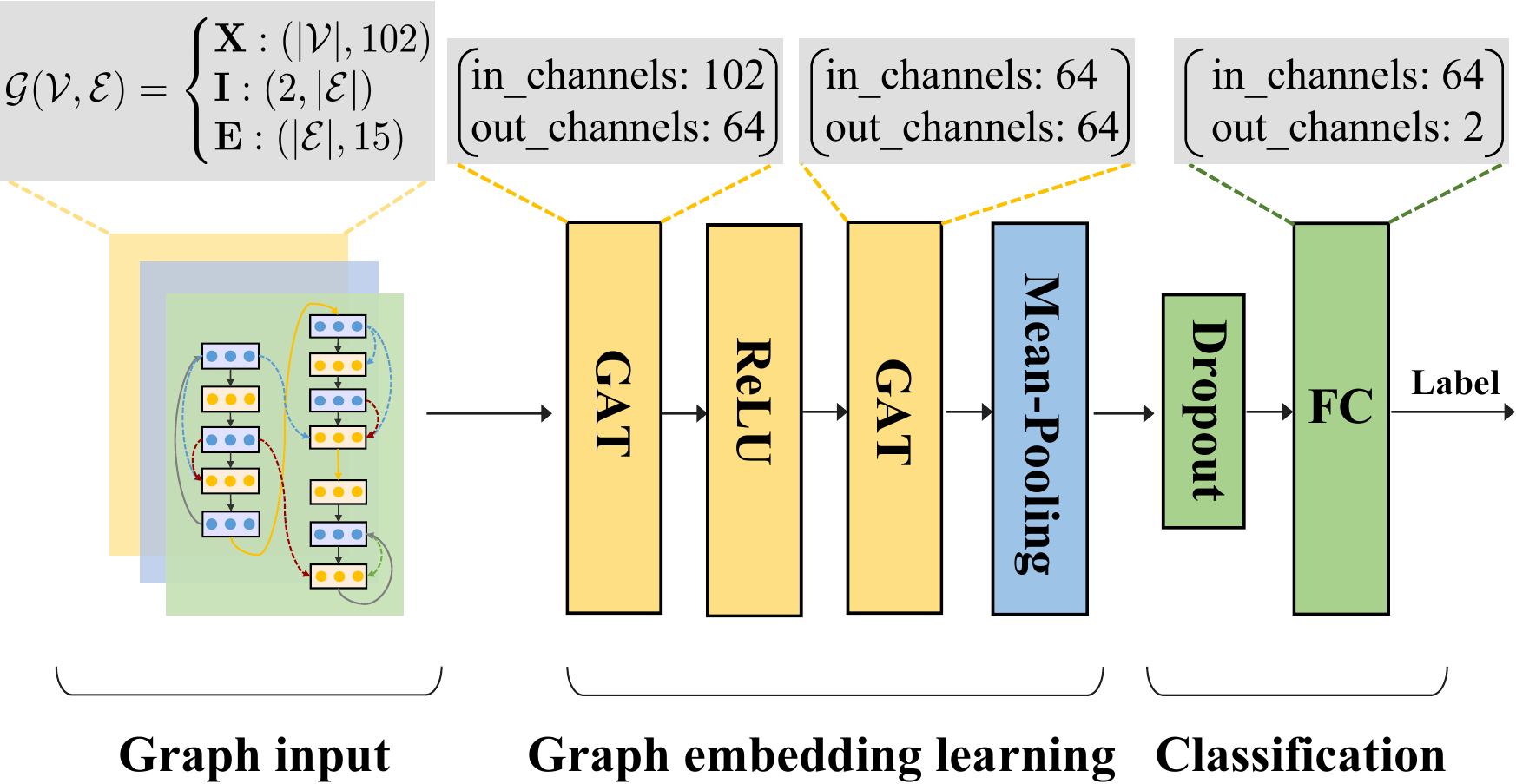}
  \caption{GNN Model.}
  \label{figure 5}
  \vspace{-15pt}
\end{figure}

Deep learning excels at automatic feature extraction from raw data and achieves top performance in many fields~\cite{CaIAuth,9552478}.
Unlike traditional deep learning models that primarily handle vector or matrix data~\cite{EchoHand}, graph neural networks (GNNs) excel at modeling and processing graph-structured data~\cite{wu2020comprehensive}.
In this section, we introduce our GNN solution to the Ponzi contract identification problem.
As illustrated in Figure~\ref{figure 5}, our GNN model consists of three parts: graph input, graph embedding learning, and classification.

\subsubsection{Graph input} 
We use CRBG ($\mathcal{G}$) as the input graph which contains nodes $\mathcal{V}=\{1, ..., n\}$ and edges $\mathcal{E}$.
The node features matrix $\textbf{X}$ has a dimension of $(|\mathcal{V}|, 102)$, where each node is represented by a 102-dimensional feature vector. 
The edge index $\textbf{I}$ has a dimension of $(2, |\mathcal{E}|)$, where each column corresponds to an edge and contains the indices of the nodes that the edge connects. 
The edge features matrix $\textbf{E}$ has a dimension of $(|\mathcal{E}|, 15)$, where each edge is represented by a 15-dimensional feature vector. 
$|\mathcal{V}|$ and $|\mathcal{E}|$ represent the number of nodes and edges in $\mathcal{G}$.

\subsubsection{Graph embedding learning} 
In graph embedding learning, we choose Graph Attention Networks (GAT) as the component of GNN convolutional layers.
GAT performs the aggregation based on the self-attention mechanism, i.e., calculating the weights between nodes and edges through learnable weight matrices $\textbf{W}$ and $\textbf{W}_e$, so that each node can be weighted and aggregated according to the characteristics of its surrounding nodes. 
Since CRBG has multi-dimensional edge features, the attention coefficients $\alpha_{i,j}$ in the self-attention mechanism are computed as:

\begin{equation}
  \alpha_{i,j} = \frac{\exp(\textbf{e}_{i,j})}{\sum_{k \in \mathcal{N}_{i\cup i}} \exp(\textbf{e}_{i,k})}
\end{equation}

\noindent
where $\textbf{e}_{i,j}$ represents the attention score indicating the importance of node $j$'s features to node $i$, and $\mathcal{N}_{i\cup i}$ represents the set of adjacent nodes of node $i$.
$\textbf{e}_{i,j}$ is obtained by concatenating the feature vectors of node $i$ and node $j$ and performing linear transformation:

\begin{equation}
  \textbf{e}_{i,j} = \text{LeakyReLU}(\vec{\textbf{a}}^T [\textbf{W} \vec{\textbf{h}}_i || \textbf{W} \vec{\textbf{h}}_j || \textbf{W}_e \vec{\textbf{m}}_{i,j}])
\end{equation}
\noindent
where $\mathrm{LeakyReLU}$ represents the activation function, $\vec{\textbf{a}}$ represents the weight vector, $||$ represents the concatenation operation, $\vec{\textbf{h}}_i$ represents the feature vector of node $i$, and $\vec{\textbf{m}}_{i,j}$ represents the multi-dimensional edge features between node $i$ and $j$.

By calculating the weight between nodes, the weighted sum of the adjacent nodes of node $i$ can be obtained:

\begin{small}
\begin{equation}
\vec{\textbf{h}}_i' = \sigma \left( \sum_{j \in \mathcal{N}_i} \alpha_{i,j} \textbf{W} \vec{\textbf{h}_j} \right)
\end{equation}
\end{small}

\noindent
where $\vec{\textbf{h}}_{i}'$ represents the updated eigenvector of node $i$, $\sigma$ represents the activation function, $\mathcal{N}_i$ represents the adjacent node of node $i$.

We set two GAT layers and use ReLU in the middle for nonlinearly transforming the node features in order to better handle the nonlinear relationship of data and increase the expressiveness of the network.
We utilize mean-pooling to aggregate the node features and obtain the global feature representation of the graph.

\subsubsection{Classification} 
The classifier comprises a dropout and a fully connected layer (FC).
The dropout randomly sets a fraction of the output of neurons to zero, which helps prevent overfitting and improves the model's generalization ability.
The purpose of a fully connected layer is to learn non-linear combinations of the features in the input data, allowing the model to make more accurate predictions.
We input the global feature representation into the classifier and obtain the predicted class label for the graph.

\section{Implementation}
\label{section Implementation}
We leverage \textsc{slither}~\cite{feist2019slither}, a static analysis framework, to extract data dependencies of smart contracts.
We instrumented the official Golang implementation of EVM (version 1.10.6)~\cite{Geth} to collect contract runtime information.
We implemented our dynamic taint engine in Golang (version 1.16.6) to cooperate with the instrumented EVM and construct the CRBG.
Our GNN model was implemented using Pytorch~\cite{pytorch}, and we employed Graph Attention Networks as the convolutional layers.

\section{Experiments}
\subsection{Research Questions}
Our test environment is comprised of a server with a 16-core Intel(R)-Xeon(R)-Gold-5218 CPU \( @\)2.30 GHz, 340GB of RAM, and the Ubuntu 18.04 LTS operating system.
We conduct experiments to answer the following four questions.

\begin{itemize}[leftmargin=4mm, itemindent=0mm]

\item \textbf{RQ1:} How effective is \textsc{PonziGuard} in identifying Ponzi contracts compared to the existing tools? 

\item \textbf{RQ2:} How does CRBG work in detecting Ponzi contracts? Is the detection process interpretable?

\item \textbf{RQ3:} How effective is the CRBG compared to the raw graph obtained directly from runtime?

\item \textbf{RQ4:} How does \textsc{PonziGuard} perform in real-world scenarios? Can it detect \textit{0-day} Ponzi schemes?

\item \textbf{RQ5:} What is the overhead of \textsc{PonziGuard}?

\end{itemize}

\renewcommand{\arraystretch}{1} 
\begin{table}[t!]
  \caption{Overall Evaluation Results. Values in parentheses represent the standard deviations across the K-fold.}
  \label{table 3}
  \centering
  \setlength{\tabcolsep}{4.8mm}{
  \begin{tabular}{c|c|c|c}
  \toprule
  \textbf{Approach} & \textbf{Precision} & \textbf{Recall} & \textbf{F1-score} \\ \midrule
  \rowcolor[HTML]{EFEFEF} 
  \textsc{OpML}\cite{DBLP:conf/www/ChenZCNZZ18}              & 89.0\% (0.05)            & 77.8\% (0.04)         & 83.0\% (0.03)           \\
  \textsc{TxML}\cite{DBLP:conf/blocksys/YuJX0X21}              & 69.6\% (0.06)             & 63.4\% (0.02)         & 66.4\% (0.07)           \\
  \rowcolor[HTML]{EFEFEF} 
  \textsc{SADPonzi}\cite{DBLP:conf/sigmetrics/ChenLSHWWL21}          & 88.1\%             & 64.5\%          & 74.5\%            \\
  \textsc{MulCas}\cite{10.1145/3571847}          & 95.1\%             & 67.4\%          & 78.9\%            \\
  \rowcolor[HTML]{EFEFEF} 
  \textsc{SourceP}\cite{10448439}          & 91.6\% (0.04)             & 93.3\% (0.03)          & 92.4\%  (0.02)           \\

  \textsc{PonziGuard}        & \textbf{96.9\%} (0.03)   & \textbf{98.2\%} (0.03) & \textbf{97.5\%} (0.02)   \\ \bottomrule
  \end{tabular}}
\end{table}

\renewcommand{\arraystretch}{1.3}

\subsection{RQ1: Effectiveness of \textsc{PonziGuard}}
\label{section Effectiveness of PonziGuard}

\subsubsection{Dataset}
\label{section Dataset}

We utilize the dataset in XBlock~\cite{PonziData} provided by Zheng et al.~\cite{10.1145/3571847}, obtained through crawling \textit{Etherscan}~\cite{Etherscan} and manual cross-checking. 
This ground-truth dataset comprises 6,498 smart contracts, with 314 identified as smart Ponzi schemes.
These smart contracts range from height 0 to height 7,500,000, and undergo a manual cross-check procedure to ensure the accuracy of their labels.

\subsubsection{Evaluation metrics}
We use the following evaluation metrics to measure the effectiveness of our approach.

Precision measures the proportion of true positive predictions made by the approach out of all positive predictions: Precision = TP / (TP + FP).
Recall measures the proportion of true positive predictions made by the approach out of all actual positive instances in the dataset: Recall = TP / (TP + FN).
F1-score is the harmonic mean of Precision and Recall, providing a single measure of the approach's overall performance: F1-score = 2 $\times$ Precision $\times$ Recall / (Precision + Recall)

\subsubsection{State-Of-The-Art}
We evaluated the effectiveness of \textsc{PonziGuard} and compared it with the studies of Chen et al.~\cite{DBLP:conf/www/ChenZCNZZ18}, Yu et al.~\cite{DBLP:conf/blocksys/YuJX0X21}, \textsc{SADPonzi}~\cite{DBLP:conf/sigmetrics/ChenLSHWWL21}, \textsc{MulCas}~\cite{10.1145/3571847} and \textsc{SourceP}~\cite{10448439}.
Chen et al.~\cite{DBLP:conf/www/ChenZCNZZ18} detect Ponzi contracts using XGBoost mainly based on the opcode frequency, and in this paper we refer to their work as \textsc{OpML}.
Yu et al.~\cite{DBLP:conf/blocksys/YuJX0X21} utilize the transactions on Ethereum to identify Ponzi contracts, and in this paper we refer to their work as \textsc{TxML}.
\textsc{SADPonzi} detects Ponzi contracts based on symbolic execution.
\textsc{MulCas} extracts contract features from multiple views and detects Ponzi schemes through multi-view training and ensemble.
\textsc{SourceP} trains the classification model by converting contract source code into Abstract Syntax Trees (AST) to extract data flow information.
\cite{DBLP:conf/www/ChenZCNZZ18,10.1145/3571847,10448439} represent state-of-the-art (SOTA) operation/source code-based machine learning approaches.
\cite{DBLP:conf/blocksys/YuJX0X21} represents SOTA transaction-based machine learning approaches, and \cite{DBLP:conf/sigmetrics/ChenLSHWWL21} represents SOTA rule-based approaches.

\subsubsection{Result and Analysis}
In our approach, we generated 6,498 graphs from 6,498 contracts in the dataset for model training and testing.
While we used the same dataset in the comparative experiment, different approaches processed the data differently.
For example, in the approach of Chen et al.~\cite{DBLP:conf/www/ChenZCNZZ18}, we compiled the 6,498 contracts into bytecode and counted the opcode frequency as inputs for model training and testing. 
In the approach of Yu et al.~\cite{DBLP:conf/blocksys/YuJX0X21}, we collected the transactions of these contracts on Ethereum and performed a random selection process to obtain a transaction network as input.
The contract bytecode could be directly applied by the symbolic execution tool of \textsc{SADPonzi}. 
For the machine learning-based approaches, we randomly divided the dataset into 5 folds and performed K-fold cross-validation.
The mean values of the evaluation metrics across the K models, as well as their corresponding standard deviations, were calculated to measure the average performances.
As shown in Table~\ref{table 3}, \textsc{PonziGuard} outperformed all the baselines on the test set, achieving 96.9\% precision, 98.2\% recall, and 97.5\% F1-score.
We believe that the poor performance of the state-of-the-art approaches can be attributed to the fact that static information cannot characterize the Ponzi contracts (\textsc{OpML}, \textsc{TxML}, \textsc{MulCas} and \textsc{SourceP}), and not all Ponzi contracts conform to the pre-defined behavior patterns (\textsc{SADPonzi}).

\begin{framed}
  \noindent
  \textbf{Answer to RQ1:} \textsc{PonziGuard} outperforms the state-of-the-art approaches in the comparative experiment, demonstrating the effectiveness of our approach in identifying Ponzi contracts.
\end{framed}

\begin{figure}[t!]
  \centering
  \includegraphics[width=3.7in]{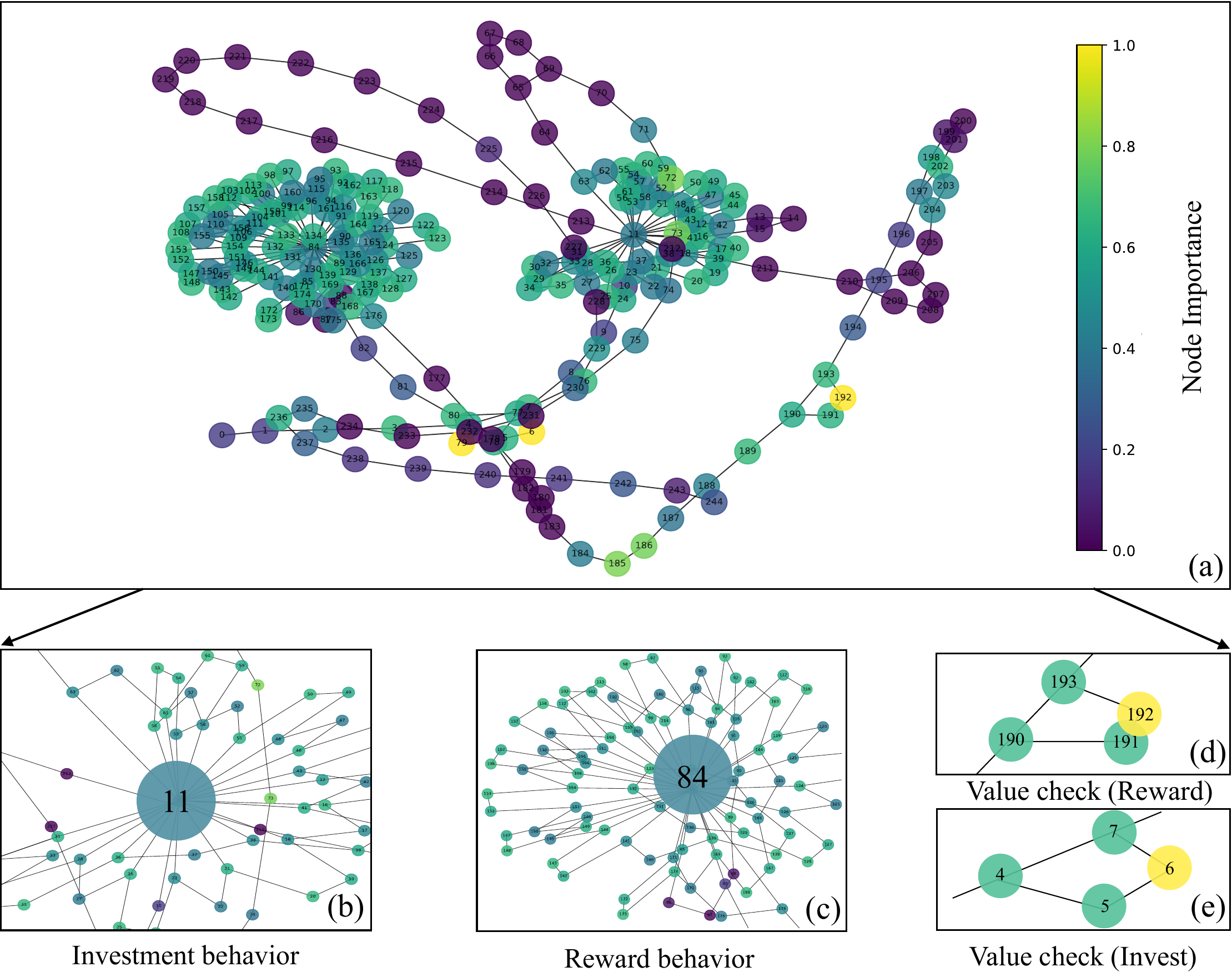}
  \caption{Node Importance Heatmap.}
  \label{figure 12}
  \vspace{-15pt}
\end{figure}

\vspace{-15pt}
\subsection{Interpretability}
\label{section Interpretability}

\begin{figure}[t!]
  \centering
  \includegraphics[width=3.7in]{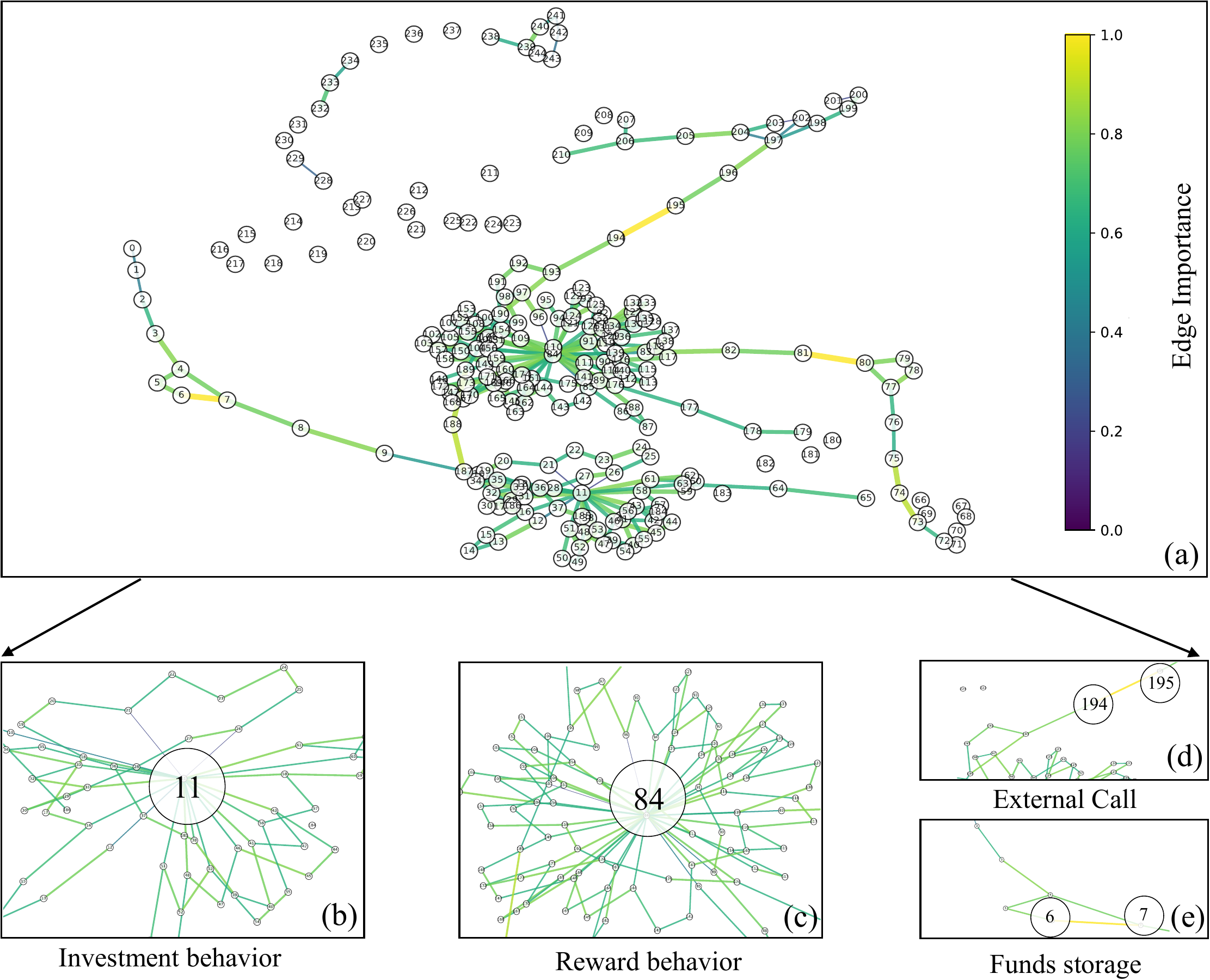}
  \caption{Edge Importance Heatmap.}
  \label{figure 13}
\end{figure}

We conduct an interpretability experiment to explain what role CRBG plays in detecting Ponzi schemes and to validate our insight into using CRBG as the key detection mechanism.
Specifically, we selected a graph from the test set as input, and then calculated the gradient of the final classification decision for each node and edge in the input graph. 
Using these gradient values, we generated importance heatmaps to highlight the nodes and edges with great impact on the classification decisions, as shown in Figure~\ref{figure 12} and Figure~\ref{figure 13}.

In Figure~\ref{figure 12}(a), some of the nodes with the greatest contribution form two clusters, and magnified details are shown in Figure~\ref{figure 12}(b) and Figure~\ref{figure 12}(c).
In these two clusters, nodes 11 and 84 are the central nodes with the most connected nodes, and correspond to the opcode \texttt{CALLVALUE} and \texttt{BALANCE}, respectively. 
These two clusters represent the investment and reward behaviors of the Ponzi scheme.
Specifically, the connections between node 11 and its surrounding nodes indicate that the contract continuously attracts new investors and records their investment amounts. 
The connections between nodes around node 84 represent the contract's cyclic operation of distributing rewards to investors.
Several other important nodes such as nodes 6 and 192 shown in Figure~\ref{figure 12}(d) and Figure~\ref{figure 12}(e), correspond to comparison operations. 
They represent the value checks before both the investment and reward behaviors.

Similarly, the most important edges in Figure~\ref{figure 13} also form two clusters, same as those observed in Figure~\ref{figure 12}, delineating the control flow and data flow of the Ponzi scheme during the investment and reward processes.
In particular, some of the most important edges such as (6, 7) and (194, 195) are presented in Figure~\ref{figure 13}(d) and Figure~\ref{figure 13}(e). 
As previously discussed, nodes 6 and 192, along with their adjacent nodes, represent the value check before the investment and reward.
Subsequently, edges (6, 7) and (194, 195) represent the contract's behaviors subsequent to the completion of these checks.
Specifically, edge (6, 7) represent the storage (\texttt{SSOTRE}) of investor's funds, while edge (194, 195) represents the external call (\texttt{CALL}) used for transferring rewards.

\begin{framed}
  \noindent
  \textbf{Answer to RQ2:} The experiment showcases that CRBG effectively reflects the characteristic behaviors of Ponzi schemes, and these characteristics greatly contribute to the classification decisions.
  This underscores CRBG's effectiveness in Ponzi scheme detection.
  Additionally, the experiment highlights that CRBG can provide valuable interpretability for our tool.  
\end{framed}

\subsection{RQ3: Effectiveness of CRBG}
In this section, we evaluate the effects of various processes applied to CRBG through a series of ablation studies, in order to determine how effective is CRBG compared to the raw graph obtained directly from runtime.

\begin{figure}[t!]
  \begin{minipage}{0.35\textwidth}
    \centering
    \includegraphics[width=\linewidth]{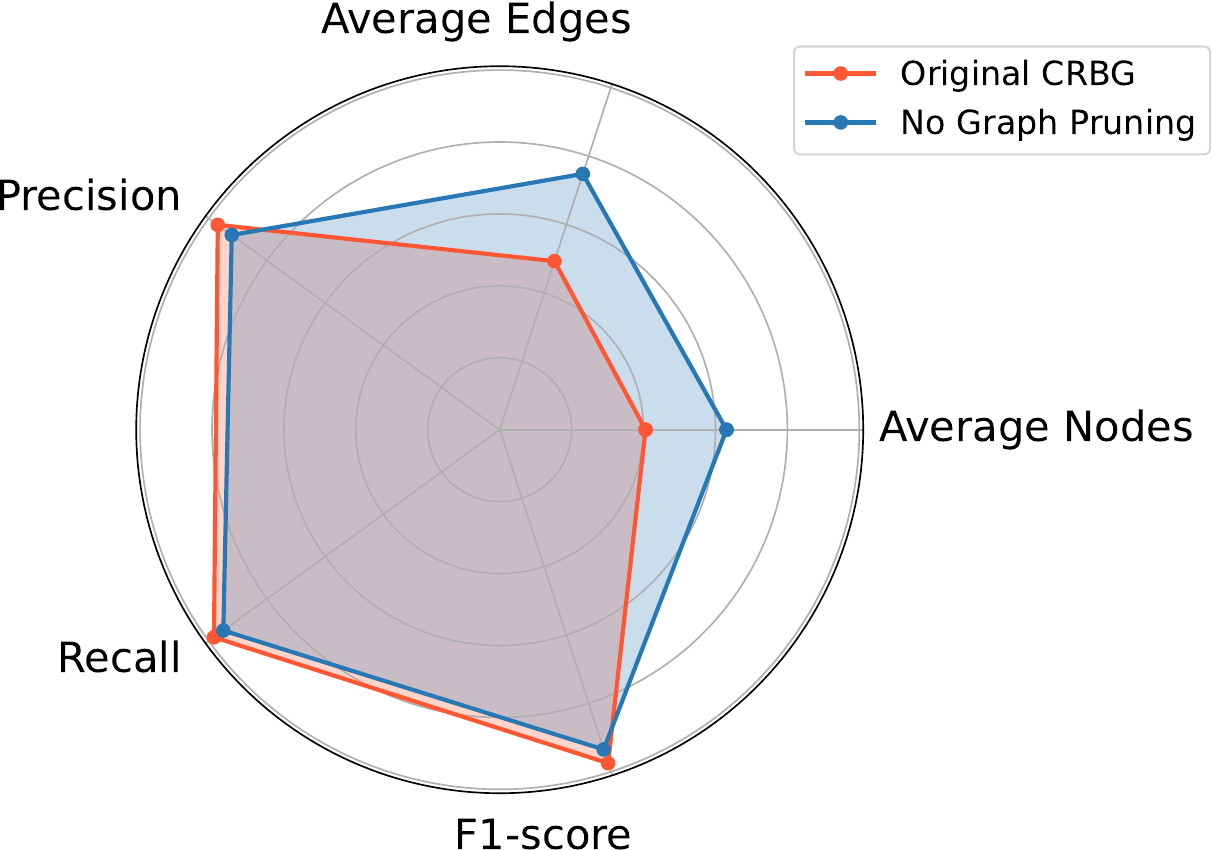}
    \caption{Comparing CRBG with ablation 1.}
    \label{figure 14}
  \end{minipage}%
  \hspace{0.05\textwidth}
  \begin{minipage}{0.35\textwidth}
    \centering
    \includegraphics[width=\linewidth]{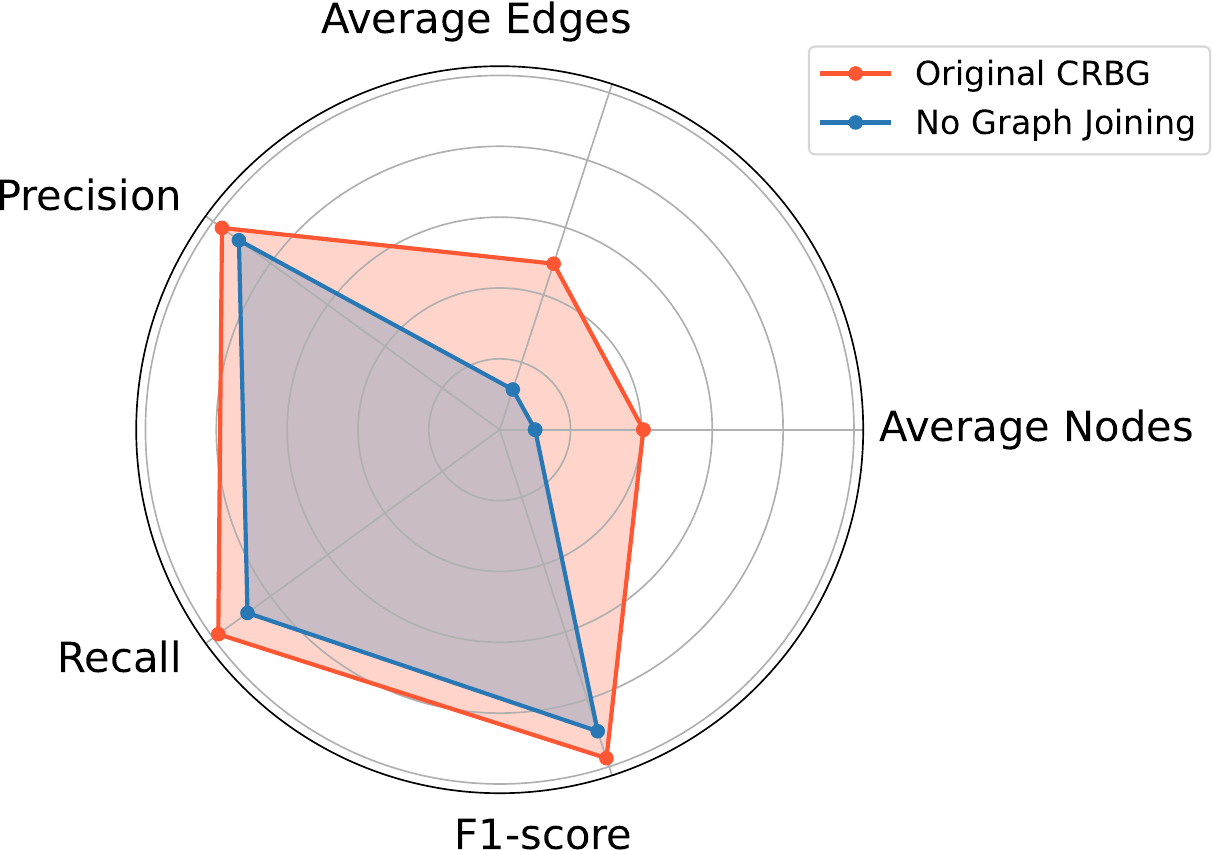}
    \caption{Comparing CRBG with ablation 2.}
    \label{figure 15}
  \end{minipage}%
\end{figure}

\begin{figure}[t!]
  \centering
  \includegraphics[width=3.4in]{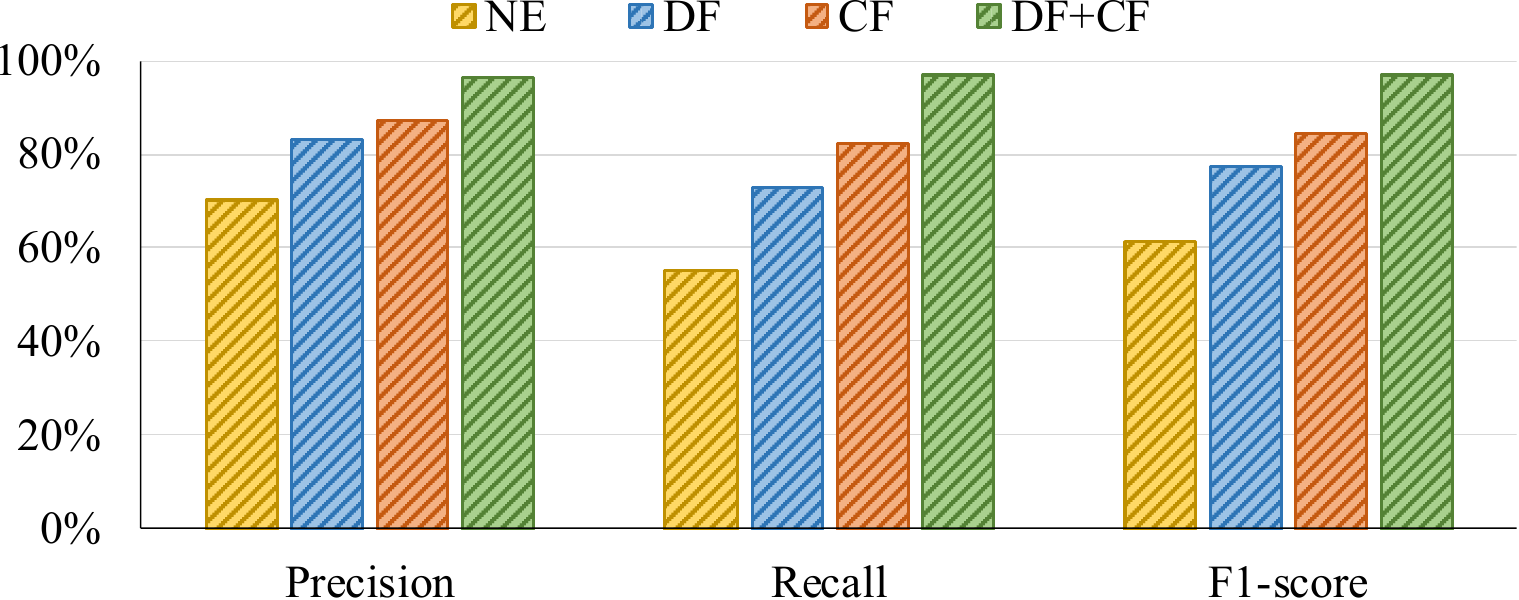}
  \caption{Performance of different ablation configurations.}
  \label{figure 6}
\end{figure}

\subsubsection{Effectiveness of graph pruning and joining}
\label{section Effectiveness of graph pruning and joining}

To evaluate the impact of graph pruning and graph joining, on the performance of CRBG, we designed two ablation settings. 
In each setting, we omitted one of these steps respectively. 
We then compared the size of CRBG generated in these two ablation settings and the corresponding experimental results on the same dataset, which is shown in Figure~\ref{figure 14} and Figure~\ref{figure 15}. 

For the CRBG without the graph pruning operation, as shown in Figure~\ref{figure 14}, the average number of nodes and edges per graph increased by nearly 30\%. 
This added significant redundant information, leading to lower precision, recall, and F1-score compared to the original CRBG. 
Additionally, the larger graph sizes increased the model's training time.
For the CRBG without the subgraph joining operation, as shown in Figure~\ref{figure 15}, the average number of nodes and edges per graph was significantly reduced to nearly one-fifth of the original CRBG. 
However, as discussed in Section~\ref{section CRBG Construction}, these smaller subgraphs failed to fully capture the contract's behavior, resulting in a decrease in performance metrics: precision, recall, and F1-score dropped by 5.9\%, 10.2\% and 8.0\%, respectively.

\subsubsection{Effectiveness of runtime data flow and control flow in CRBG}
To gain a better understanding of the effectiveness of control and data flow in CRBG, we performed an ablation study by configuring \textsc{PonziGuard} in four distinct modes: data flow only (\textbf{DF}), control flow only (\textbf{CF}), both (\textbf{DF+CF}), and neither (\textbf{NE}).
In \textbf{DF} mode, we removed control flow edges and only kept data flow edges in CRBG. 
On the contrary, in \textbf{CF} mode, we removed data flow edges and only kept control flow edges.
In \textbf{NE} mode, we removed both data flow edges and control flow edges in CRBG.
The mode \textbf{DF+CF} is the native \textsc{PonziGuard} which includes both control and data flow.

Figure~\ref{figure 6} shows the performance of these four modes after 5-fold cross-validation on the same dataset.
Compared to the native \textsc{PonziGuard} baseline, there were drops of 9.3\%, 15\%, and 12.2\% in the evaluation metrics of the \textbf{CF} mode.
In the \textbf{DF} mode, the evaluation metrics decreased to a greater extent compared to the native \textsc{PonziGuard} baseline, with a drop of 13.5\%, 24.2\%, and 19.3\%, respectively.
Undoubtedly, \textbf{NE} mode exhibited the worst performance, with a significant drop of 26.4\%, 42.1\%, and 35.3\%, respectively.
The main reason for the poor performance of the \textbf{CF} mode is that, with control flow only, \textsc{PonziGuard} cannot capture the flow of investors' investments in contracts. 
Therefore, some contracts with Ether redistribution logic may be misreported. 
On the other hand, the lack of control flow in the \textbf{DF} mode results in the loss of contract context information, such as the functions and order in which variables are used.
This ablation study highlights that both control flow and data flow are crucial in capturing the behavioral patterns of Ponzi contracts, and the gathering of this runtime information significantly improves the performance of \textsc{PonziGuard}.

\begin{table}[t!]
  \caption{Comparing model performance between different node embedding settings.}
  \label{table 5}
  \centering
  \begin{tabular}{c|c|c|c|c|c}
  \toprule
  \textbf{Test} & \textbf{\begin{tabular}[c]{@{}c@{}}Node Embeddings adopted in Test Set\end{tabular}} & \textbf{Model} & \textbf{Precision} & \textbf{Recall} & \textbf{F1-score} \\ \midrule
  \rowcolor[HTML]{EFEFEF} 
  1             & One-hot vectors                                                                         & MOE            & 88.5\%              & 82.1\%           & 85.2\%             \\
  2             & Enhanced                                                            & MEE            & 96.4\%              & 96.4\%           & 96.4\%             \\
  \rowcolor[HTML]{EFEFEF} 
  3             & Variant 1                                                                               & MEE            & 96.3\%              & 92.9\%           & 94.5\%             \\
  4             & Variant 2                                                                               & MEE            & 96.2\%              & 89.3\%           & 92.6\%             \\ \bottomrule
  \end{tabular}
  \end{table}

\subsubsection{Effectiveness of node embeddings adopted in CRBG}
In Section~\ref{section CRBG Construction}, we utilized Doc2Vec to enhance node embeddings in CRBG based on the operation descriptions from the Ethereum Yellow Paper.
We believe that the semantic information conveyed in these descriptions is representative and can capture the relationships between the nodes in CRBG.
We conducted a comparative experiment to evaluate the efficacy of the node embeddings we enhanced.
In this comparative experiment, one model was trained on the dataset described in Section~\ref{section Dataset} using our enhanced node embeddings (Model with enhanced node embeddings, abbreviated as \textbf{MEE}).
In contrast, another model was trained on the same dataset, but replacing our node embeddings with one-hot vectors (Model with one-hot vectors as node embeddings, abbreviated as \textbf{MOE}).
We used 80\% of the dataset for training and 20\% for testing. 
As shown in Table~\ref{table 5}, compared with the one-hot vectors as node embeddings (Test 1), the node embeddings based on the operation description (Test 2) performed better in the evaluation metrics.
It demonstrates that the node embeddings generated from operation descriptions capture the underlying semantics, leading to a better understanding of the graph's structure and properties, which accounts for this performance improvement.

\begin{table}[t!]
  \caption{Variants of opcode description.}
  \label{table 6}
  \setlength{\tabcolsep}{3.4mm}{
  \begin{tabular}{c|c|c}
  \hline
  \textbf{Value}        & \textbf{Mnemonic}      & \textbf{Original Description:}                               \\ \cline{1-2}
  \multirow{5}{*}{0x57} & \multirow{5}{*}{JUMPI} & \textit{"Conditionally alter the program counter."}          \\ \cline{3-3} 
                              &                        & \textbf{Synonyms substitution:}                              \\
                              &                        & \textit{"Conditionally change the instruction pointer."}     \\ \cline{3-3} 
                              &                        & \textbf{Changing sentence structure or grammar:}             \\
                              &                        & \textit{"Alter the program counter based on the condition."} \\ \hline
  \end{tabular}}
\end{table}

To ascertain that the performance improvement is primarily attributed to the semantics themselves, rather than the way the semantics are described, we conducted additional analysis.
We rewrote the operation descriptions in the Ethereum Yellow Paper with two principles while retaining the original semantics.
The first principle is using synonyms substitution.
By substituting words with their synonyms, we retained the original semantics while using a different expression.
Another principle is changing sentence structure or grammar.
This can be done by using different sentence patterns, altering the word order, or adjusting the placement of clauses.
For instance, as shown in Table~\ref{table 6}, the description for \texttt{JUMPI} in Table~\ref{table 2} can be rewritten as "\textit{Conditionally change the instruction pointer}" and "\textit{Alter the program counter based on the condition}" according to these two principles.
Based on these two alternative description rewriting principles, we re-extracted the node embeddings and created two variants of the test set (Variant 1 and Variant 2).
Then, we evaluated our model (\textbf{MEE}, the model trained with enhanced node embeddings) on these two variant test sets.
As shown in Table~\ref{table 5}, our model exhibited similar performance on the variant test sets (Test 3 and 4) compared to the native test set (Test 2), suggesting that the improvement in model performance is primarily attributed to the semantic information rather than the specific way in which the semantics are conveyed.

\begin{framed}
  \noindent
  \textbf{Answer to RQ3:} CRBG proves effective compared to the raw graph obtained directly from runtime. And ablation studies further confirm that our processes on CRBG such as graph pruning, subgraph joining, and enhanced node embeddings significantly enhance detection performance.
\end{framed}

\subsection{RQ4: Performance in real-world scenarios}
\label{section Performance in real-world scenarios}

To evaluate the effectiveness of \textsc{PonziGuard} in real-world scenarios, we conducted two experiments on the Ethereum Mainnet. 
The first is a large-scale transaction replay on the Ethereum historical blocks, to assess the number and economic impact of Ponzi contracts over the past few years. 
The second involved collecting recently deployed contracts to detect \textit{0-day} Ponzi schemes.

\subsubsection{Historical Transaction Replay}
Firstly, we ran the \textit{Geth} client with the option: \textit{sync-mode-full} to synchronize with the Ethereum Mainnet. 
The number of smart contracts has experienced explosive growth in recent years (about one million per quarter~\cite{ethereum-statistics}), which is a significant amount for our approach based on runtime information.
Therefore, we set the synchronization time until January 2022 (approximately 14,000,000 blocks), only as a preliminary experiment to verify the performance of \textsc{PonziGuard} in real-world scenarios.
Then, we replaced the native EVM with our instrumented EVM and integrated the dynamic taint engine.
We re-executed every transaction on the synchronized blockchain from the genesis block, which is a time-consuming process.
Finally, we fed the generated graphs into our GNN model for prediction.
As a result, \textsc{PonziGuard} successfully identified 805 Ponzi contracts on Ethereum Mainnet, out of which 497 contracts have accessible source code on \textit{Etherscan}~\cite{Etherscan}.
We randomly selected 50 contracts\footnote{\url{https://github.com/PonziDetection/PonziGuard/tree/main/dataset/Result/verified}} from these 497 contracts and conducted a manual examination through Remix~\cite{remix}, a solidity IDE, to ensure they meet our predefined criteria for Ponzi contracts, resulting in a 100\% true positive rate.
To gain deeper insights into these 805 Ponzi contracts, we conducted further analysis using the data collected from Etherscan in the remainder of this section.

\begin{figure*}[t!]
  \begin{minipage}{0.42\textwidth}
    \centering
    \includegraphics[width=\linewidth]{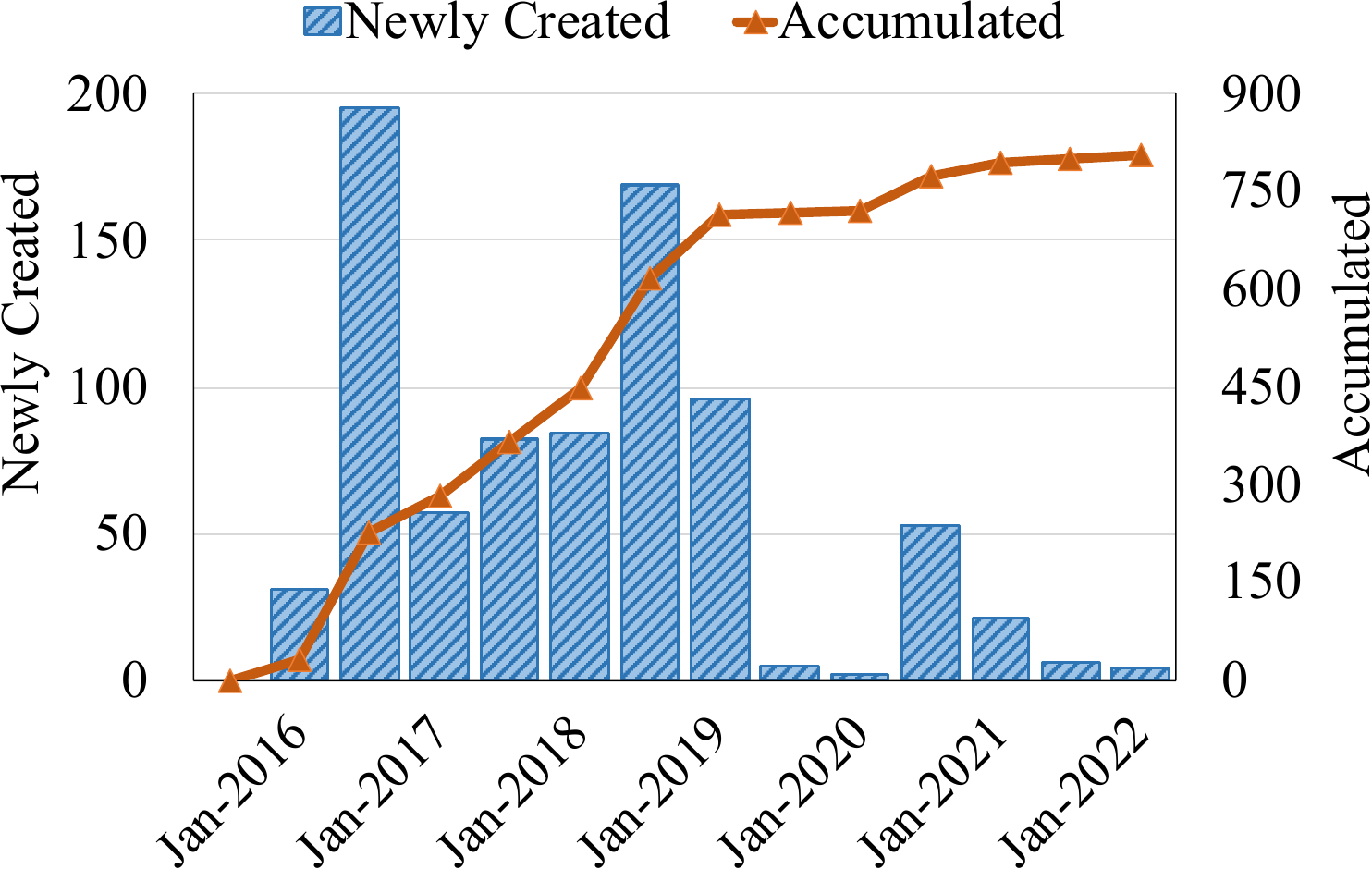}
    \caption{Distribution of Ponzi schemes creation.}
    \label{figure 7}
  \end{minipage}%
  \hspace{0.05\textwidth}
  \begin{minipage}{0.42\textwidth}
    \centering
    \includegraphics[width=\linewidth]{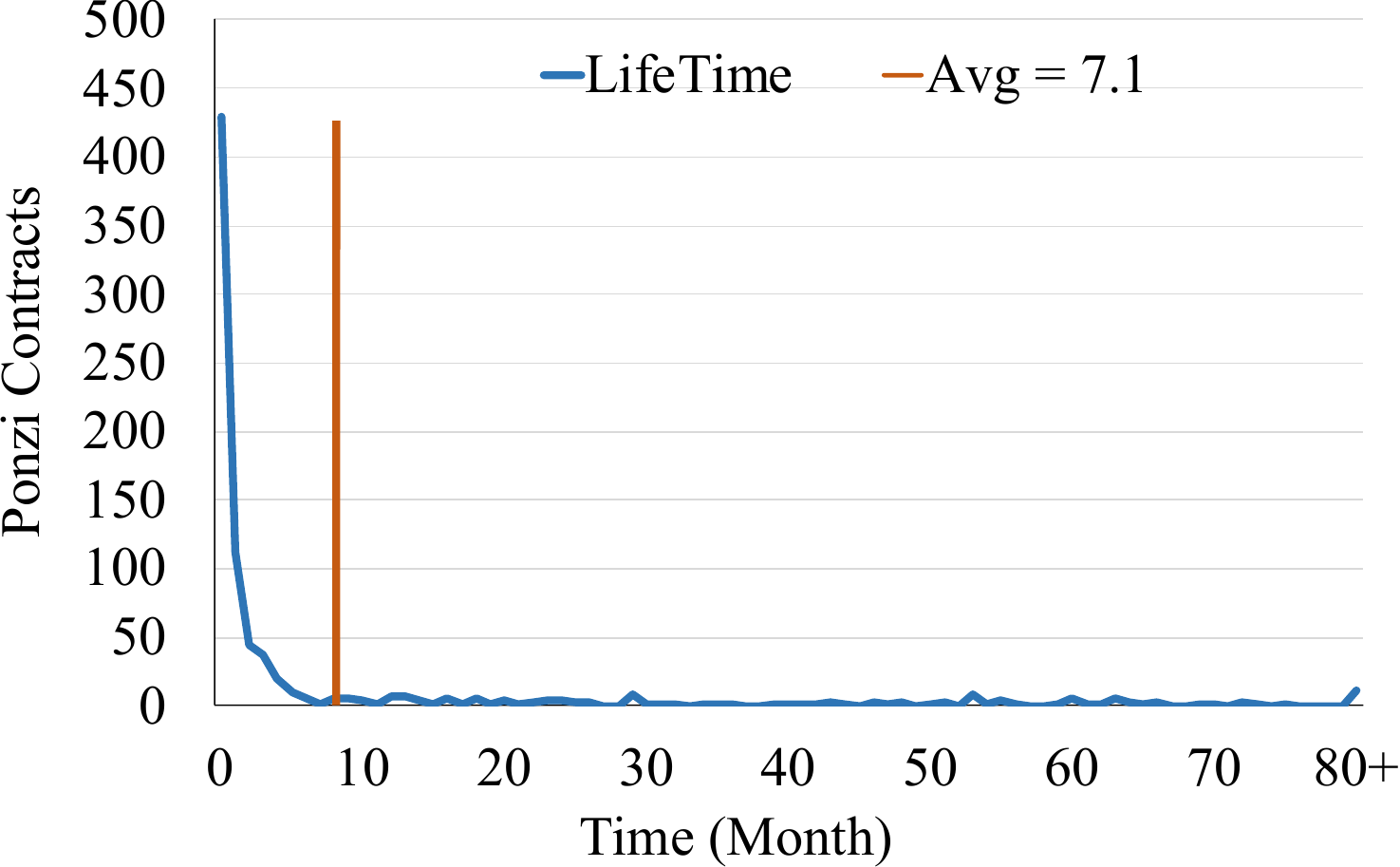}
    \caption{Lifetime/Average Lifetime of Ponzi schemes.}
    \label{figure 8}
  \end{minipage}%
\end{figure*}

\begin{figure*}[t!]
  \begin{minipage}{0.42\textwidth}
    \centering
    \includegraphics[width=\linewidth]{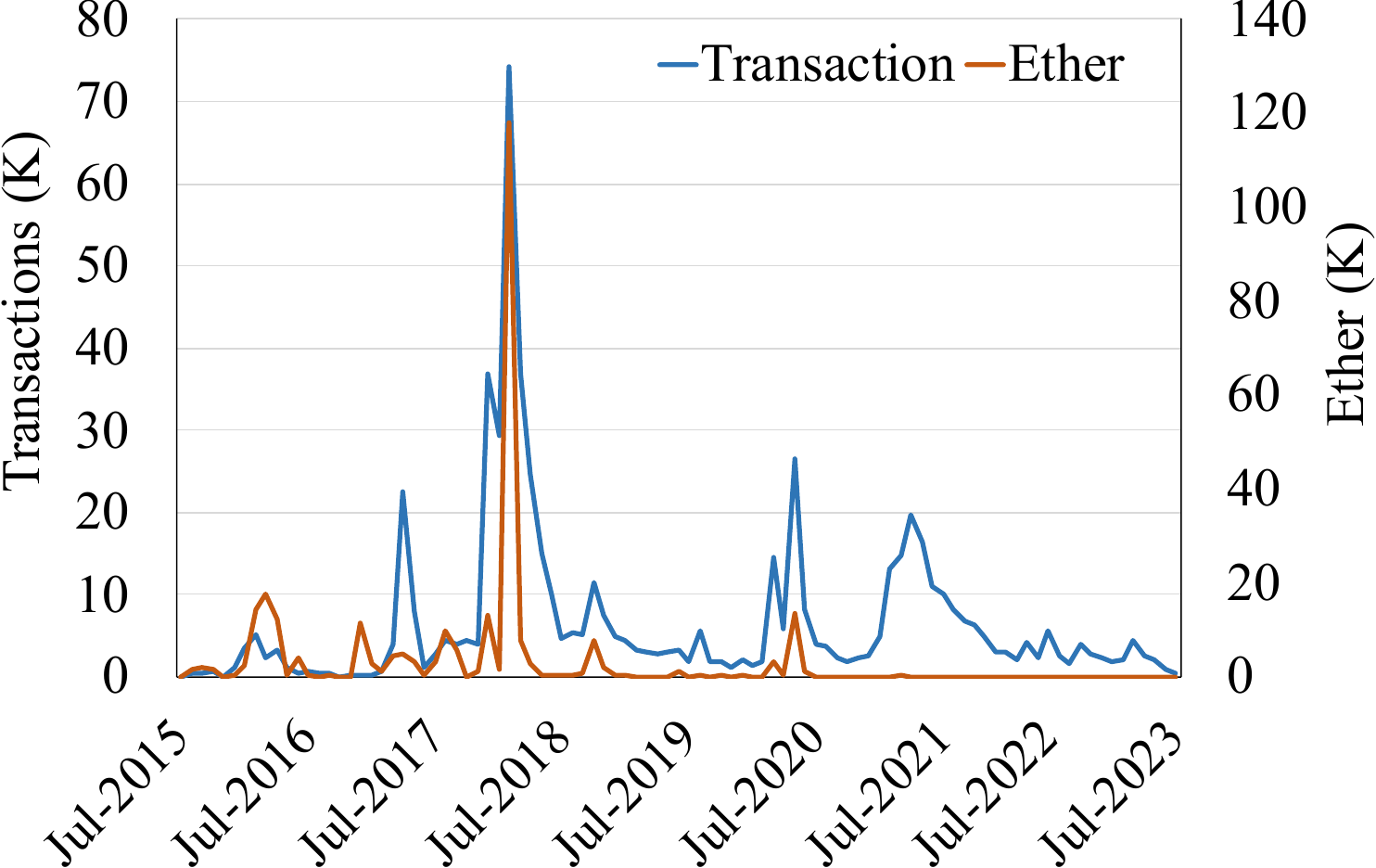}
    \caption{Transaction flow and Ether flow of Ponzi schemes over time.}
    \label{figure 9}
  \end{minipage}%
  \hspace{0.05\textwidth}
  \begin{minipage}{0.42\textwidth}
    \centering
    \includegraphics[width=\linewidth]{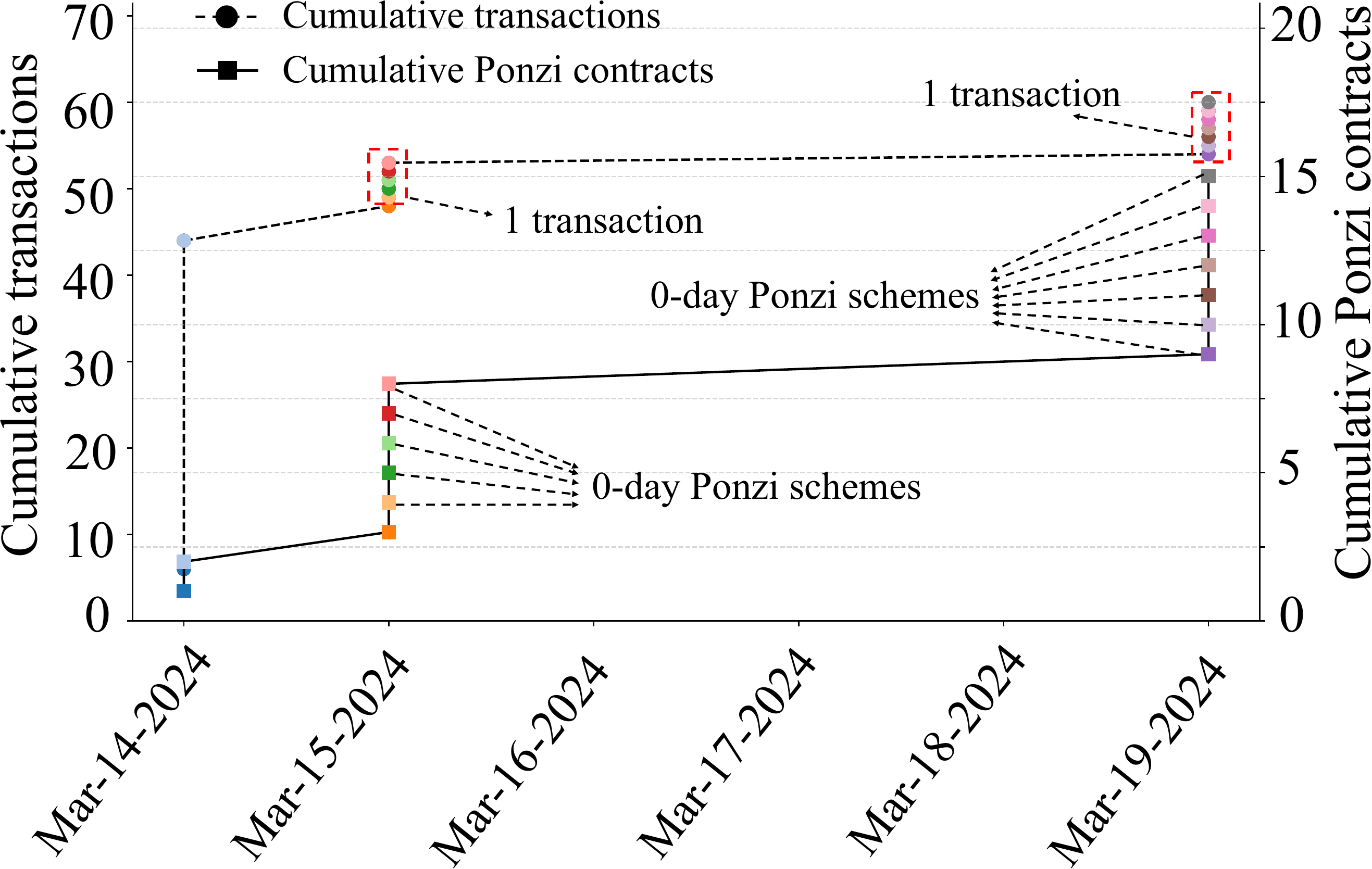}
    \caption{\textit{0-day} Ponzi schemes in 10,000 recently deployed smart contracts.}
    \label{figure 16}
  \end{minipage}%
\end{figure*}

\textbf{Creation Time of Ponzi Contracts}.
Figure~\ref{figure 7} shows the distribution of these 805 Ponzi contracts.
Ponzi schemes started appearing on Ethereum as early as 2015. 
Subsequently, the rapid development of Ethereum led to a significant growth of Ponzi contracts during the years 2016-2019. 
Then, we witnessed a brief recession in Ponzi schemes possibly linked to the impact of the COVID-19 pandemic~\cite{ponzitracker}.
The global crypto mining boom in 2021~\cite{crypto-boom} resulted in another minor peak in Ponzi schemes.
With the increasing popularity of various tokens on Ethereum, ERC-20 Tokens for instance, we anticipate another peak in Ponzi schemes on Ethereum in the near future.

\textbf{Lifetime of Ponzi Contracts}
We regard the time from the creation of a Ponzi contract to its last transaction as its lifetime.
We investigated the lifetime of these 805 Ponzi contracts, as shown in Figure~\ref{figure 8}.
While some of these contracts remain active in 2023\footnote{For instance: 0xa90be2201bfed97587a2a17949e8624eafe51d13 and 0xf8f04b23dace12841343ecf0e06124354515cc42}, the majority of Ponzi contracts have a lifetime of less than three months, and their average lifetime is about seven months.
As for the short lifetime of the Ponzi, some ended because the scam failed to attract new investors, resulting in its collapse. 
Others were ended because the scam owner intentionally triggered the self-destruct function of the contracts and absconded with the funds.
These findings indicate that Ponzi contracts are likely to collapse within a short period of time, and most of their users will not be unable to reclaim their promised returns.

\textbf{Financial Impact}
We analyzed the financial impact of the 805 Ponzi contracts identified by \textsc{PonziGuard} on Ethereum Mainnet by aggregating their transactions and the inflow of Ether.
Figure~\ref{figure 9} shows their monthly distribution, revealing a positive correlation between the inflow of Ether and the number of transactions of the contracts. 
The peak was in February 2018, when a total of 117,953 Ether flowed into Ponzi contracts, equivalent to \$108 million at the exchange rate of that time. 
From January 2015 to July 2023, 615,483 transactions, totaling 281,700 Ether, flowed into Ponzi contracts.
At the current exchange rate, the value of these tokens can reach as high as \$500 million.
It is also evident that, in recent years, the involvement of Ether may not be substantial in a Ponzi scheme, as some of them began adopting ERC tokens for investments and rewards.
However, it is important to note that such kinds of Ponzi contracts still meet the criteria outlined in Section~\ref{section Ponzi Schemes}, and our method remains effective in identifying them\footnote{Evidenced by the example of 0xb3836d31d43d315ba74c21aad3818f9378256152}.

\subsubsection{Detecting 0-day Ponzi Schemes}
\label{section Detecting 0-day Ponzi Schemes}

To ascertain whether new Ponzi schemes still keep emerging and whether our tool can detect them effectively, we collected the source code of about 10,000 smart contracts recently deployed in March 2024 from Etherescan~\cite{Etherscan} and employed \textsc{PonziGuard} for detection. 
After manual confirmation, we identified 15 Ponzi schemes among these 10,000 recently deployed contracts, and their address are available at our GitHub repository\footnote{\url{https://github.com/PonziDetection/PonziGuard/tree/main/dataset/Result}}.
As shown in Figure~\ref{figure 16}, most of these Ponzi contracts were newly deployed and have only one transaction (creation transaction) on them, showcasing the ability of \textsc{PonziGuard} to uncover 0-day Ponzi schemes.

\begin{framed}
  \noindent
  \textbf{Answer to RQ4:} \textsc{PonziGuard} successfully identified 805 Ponzi contracts in the Ethereum historical blocks and found 15 latest 0-day Ponzi contracts deployed within a month, demonstrating the effectiveness of \textsc{PonziGuard} in real-world scenarios.
  These contracts have resulted in significant financial losses, amounting to millions of USD, which emphasizes the severity of Ponzi contracts on Ethereum and the urgency of identifying them effectively.
\end{framed}

\subsection{RQ5: Overhead of \textsc{PonziGuard}}
In \textsc{PonziGuard}, we instrument the EVM and build a dynamic taint engine to obtain contract runtime information, which introduces a certain amount of time overhead compared to the native smart contract execution environment.
We conducted experiments to evaluate this overhead.

\subsubsection{Ground-Truth Dataset}
\begin{figure}[t!]
  \begin{minipage}{0.42\textwidth}
    \centering
    \includegraphics[width=\linewidth]{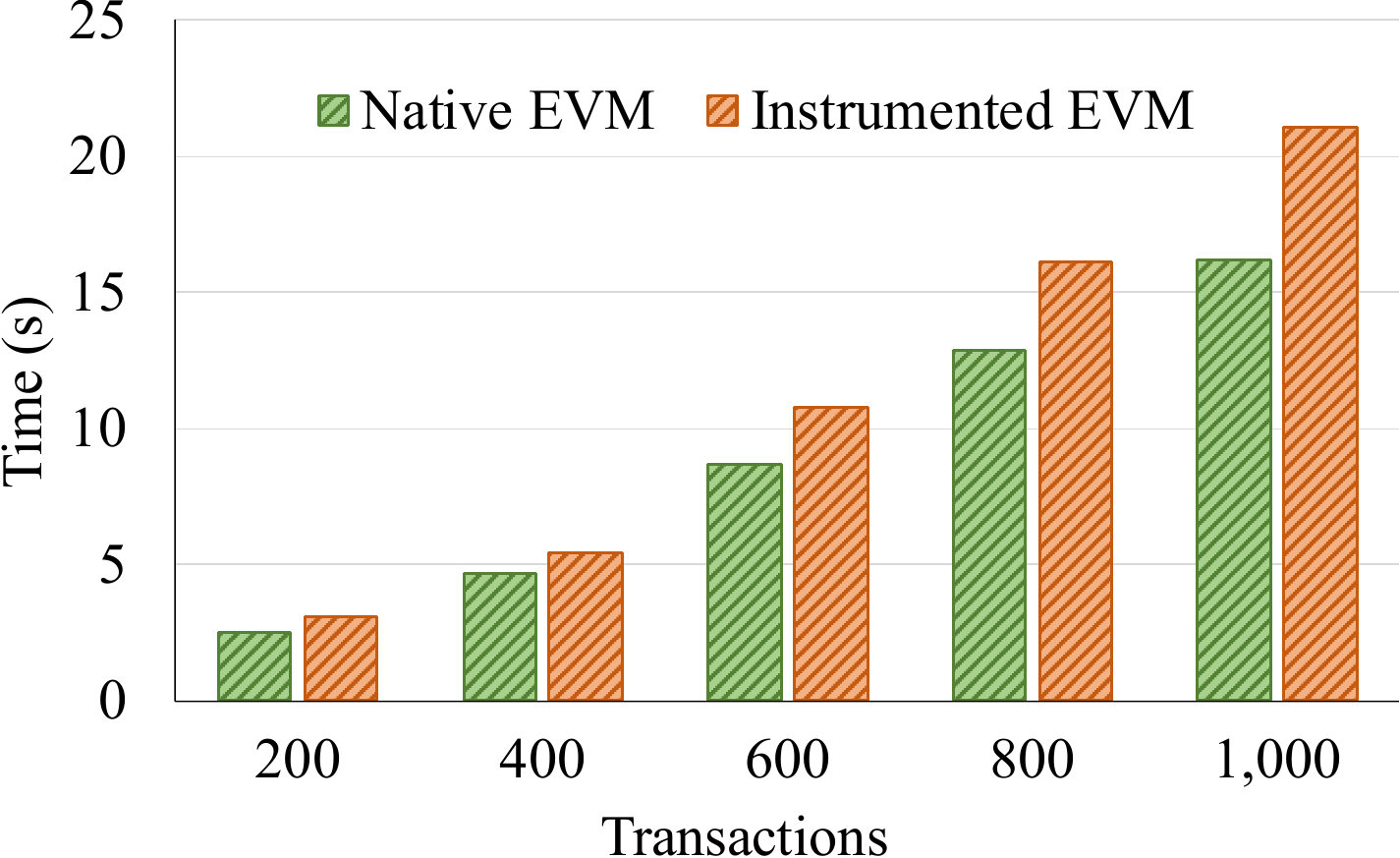}
    \caption{Overhead on the ground-truth dataset.}
    \label{figure 10}
  \end{minipage}%
  \hspace{0.05\textwidth}
  \begin{minipage}{0.42\textwidth}
    \centering
    \includegraphics[width=\linewidth]{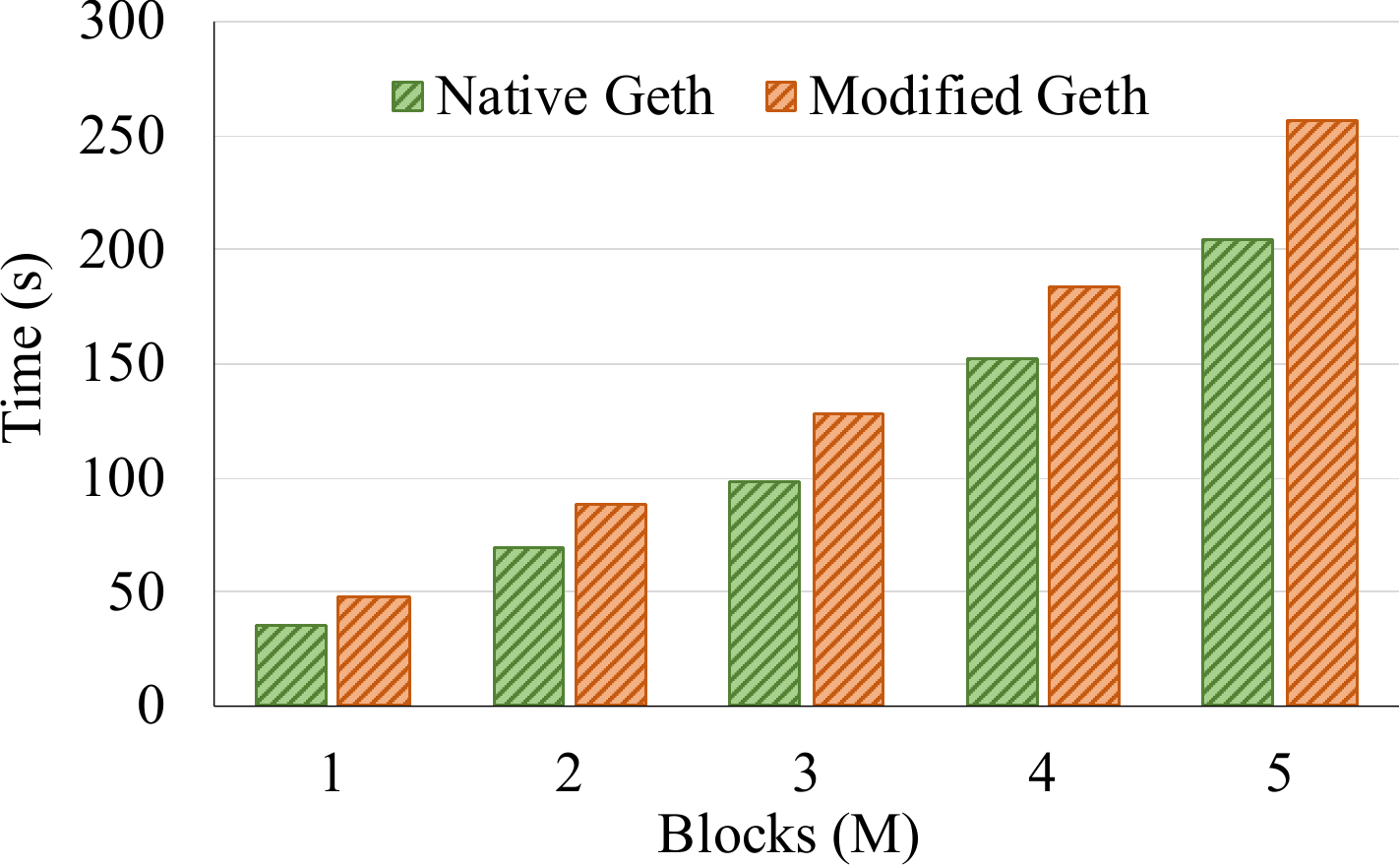}
    \caption{Overhead in real-world scenarios.}
    \label{figure 11}
  \end{minipage}%
\end{figure}

For the experiment described in Section~\ref{section Effectiveness of PonziGuard}, the contracts were executed in an independent instrumented EVM with the taint engine. 
To evaluate the time overhead, we generated 1,000 transactions for a contract and sent them along with the contract to both the native EVM and the instrumented EVM separately.
To accurately assess the time overhead, we repeated the process 10 times and recorded the average time it took to process these transactions. 
The results, shown in Figure~\ref{figure 10}, indicate that when the processing of 1,000 transactions was completed, the average overhead of the instrumented EVM reached a maximum of approximately 30.2\%.

\subsubsection{Real-World Scenarios}
In the experiment described in Section~\ref{section Performance in real-world scenarios}, we conducted re-execution of historical transactions on the synchronized blockchain.
To evaluate the time overhead, we re-executed the transactions of the first 500,000 blocks on the synchronized blockchain using both the native \textit{Geth} and the \textit{Geth} modified by \textsc{PonziGuard} separately. 
To accurately assess the time overhead, we repeated the process 10 times and recorded the average time it took to complete the re-execution.
As shown in Figure~\ref{figure 11}, when it comes to 500,000 blocks, the time overhead amounts to 25.5\%, which is smaller than the overhead on the ground-truth dataset.
This can be attributed to the fact that re-execution involves additional reading and verification operations on the blockchain, in addition to the time consumed by contract execution.

\begin{framed}
  \noindent
  \textbf{Answer to RQ5:} The time overhead introduced by our taint engine and the modification of EVM is an acceptable compromise to obtain contract runtime information.
\end{framed}

\section{Related Work}
In this section, we first describe the previous studies about Ponzi schemes on Ethereum. 
Then, we describe the studies related to the techniques we use.
\subsection{Ponzi Scheme on Ethereum}
Bartoletti et al.~\cite{DBLP:journals/fgcs/BartolettiCCS20}, the first to study Ponzi schemes on Ethereum, use the Normalized Levenshtein Distance (NLD) to measure the similarity of contract bytecode.
Similarly, the rule-based approaches have been developed by Sun et al.~\cite{DBLP:conf/qrs/SunXY020} who leverage behavior forest similarity to detect Ponzi contracts, and Chen et al.~\cite{DBLP:conf/sigmetrics/ChenLSHWWL21} who use symbolic execution for detection.
These approaches require a comprehensive summary of existing Ponzi schemes and expert experience.
However, it is challenging to cover all possible scenarios based on the existing known Ponzi contracts, which limits their capability to detect Ponzi contracts that fall outside the scope of the summarized rules.
Additionally, other approaches~\cite{DBLP:conf/www/ChenZCNZZ18,DBLP:conf/blockchain2/JungTGG19,DBLP:conf/ijcnn/FanFXZ20,DBLP:conf/IEEEscc/LouZC20,DBLP:conf/blocksys/YuJX0X21,10448439,10.1145/3571847} use static information like opcode or transactions for machine learning models to improve detection capabilities.
However, these approaches suffer from the limitation that static information cannot well distinguish Ponzi contracts from other contracts, and transaction-based machine learning approaches cannot detect \textit{0-day} Ponzi schemes.

\subsection{Smart Contract Fuzzing}
Fuzzing has been proven to be effective to exploit vulnerabilities in smart contracts~\cite{9842653,DBLP:conf/kbse/0001LC18,DBLP:conf/ccs/HeBATV19,DBLP:conf/sigsoft/WustholzC20,DBLP:conf/eurosp/TorresIGS21,DBLP:conf/icse/NguyenP0L020,DBLP:conf/wcre/ZhangWLM20}.
ContractFuzzer~\cite{DBLP:conf/kbse/0001LC18} is a black-box fuzzer for Ethereum smart contracts to detect security bugs such as gasless send and timestamp dependency.
Some grey-box fuzzers~\cite{DBLP:conf/ccs/HeBATV19,DBLP:conf/sigsoft/WustholzC20,DBLP:conf/eurosp/TorresIGS21,DBLP:conf/icse/NguyenP0L020,DBLP:conf/wcre/ZhangWLM20,xFuzz,ItyFuzz,10.1145/3597926.3598057} have also been proposed for smart contracts.
These methods are designed to exploit vulnerabilities in smart contracts, while \textsc{PonziGuard} uses fuzzing to invoke contracts and obtain their runtime information.

\subsection{Taint Analysis}
Taint analysis is an effective technique to analyze the data flow in programs~\cite{10.1145/3589334.3645579,10172538,9726808}.
There have been studies that leverage taint analysis to help analyze smart contract such as Osiris~\cite{10.1145/3274694.3274737}, Sereum~\cite{DBLP:conf/ndss/RodlerLKD19} and EthPloit~\cite{DBLP:conf/wcre/ZhangWLM20}.
Osiris~\cite{10.1145/3274694.3274737} is an integer bug detection framework that combines taint analysis and symbolic execution.
Sereum~\cite{DBLP:conf/ndss/RodlerLKD19} leverages taint analysis to protect smart contracts with re-entrancy vulnerabilities from being exploited.
EthPloit~\cite{DBLP:conf/wcre/ZhangWLM20} adopts taint analysis to generate exploit-targeted transaction sequences, in order to make the contract fuzzing process more efficient.
Those studies are orthogonal to this paper: they aim to uncover security vulnerabilities in smart contracts, while our tool is designed specifically for for identifying malicious contracts.

\subsection{Graph Neural Network}
Graph Neural Networks (GNNs) are a subset of deep learning techniques that have shown remarkable effectiveness across various domains, including user authentication~\cite{251562,10458004} and mitigating vulnerabilities~\cite{10.1145/3429444}.
GNNs are designed to process and learn from data that is structured in the form of graphs~\cite{wu2020comprehensive}.
They have been shown highly effective in various tasks, such as node classification~\cite{ivanov2021boost,DBLP:conf/iclr/ThakoorTAADMVV22}, link prediction~\cite{zhang2021labeling,you2019position}, and graph classification~\cite{bouritsas2022improving,10.1145/3584945}. 
In this paper, we leverage GNNs for CRBG analysis and formulate the detection of Ponzi contracts as a graph classification task.

\begin{lstlisting}[language = Solidity, escapechar=!, caption = Snippet of squareRootPonzi, label = lst 2, float=t]
  !\textbf{Origin}!:
  uint index = Calculator.length + 1;
  Calculator[index].ethereumAddress = msg.sender;
  Calculator[index].name = "masterly calculated";
  

  !\textbf{Modified}!:
  !\colorbox{light-gray}{uint index = Calculator.length;}!
  !\colorbox{light-gray}{Calculator.length += 1;}!
  Calculator[index].ethereumAddress = msg.sender;
  Calculator[index].name = "masterly calculated";
  \end{lstlisting}

\section{Discussion}
We note that some static analysis tools~\cite{feist2019slither,schneidewind2020ethor,Mythril,Chen2020SODAAG} can obtain the \textit{Static} control and data flow with lower overhead, which also reflect the contract behavior to some extent.
In this section, we provide the explanation for our decision to use \textit{Runtime} information rather than \textit{Static} information to construct our CRBG.

Firstly, static analysis is inherently imprecise following the principle of over-approximation.
This conservative approach preserves all "could happen" or "could exist" cases, which is useful for capturing program errors and vulnerabilities but inappropriate for characterizing a program's behavior.
For instance, squareRootPonzi\footnote{0x8ea6c8077d6316b46e449aec8fb60a606cf50eea} is a false positive case in the previous study~\cite{DBLP:journals/fgcs/BartolettiCCS20}, and its code snippet is shown in Listing~\ref{lst 2}.
This contract appears to follow the logic of a typical Ponzi scheme, however, the incorrect assignment to the variable \texttt{index} will cause the typical \textit{IndexError} during its runtime.
Consequently, the contract will always exit with an error.
The correct code is demonstrated in Line 8 and Line 9.
However, static analysis tools cannot recognize this invalid execution path due to the lack of runtime information, and following the principle of over-approximation.
If we utilize the static information to characterize the contract behaviors, it is likely to misreport it as a Ponzi scheme.

Secondly, the output of static analysis includes all possible execution paths and data flows of the contract, making it challenging to determine which information should be pruned.
Constructing this information into a graph structure can result in a significant increase in size and contain redundant data, which is not efficient for model training.

\section{Conclusion}
In this paper, we propose \textsc{PonziGuard}, an approach for identifying Ponzi schemes on Ethereum based on the \textit{contract runtime behavior graphs} (CRBG).
The experimental results demonstrate that \textsc{PonziGuard} is effective on both the ground-truth dataset and real-world scenarios with acceptable overhead. 
Moreover, our preliminary experiment conducted on Ethereum Mainnet has identified 805 Ponzi contracts that have caused millions of USD in financial losses.
We also found \textit{0-day} Ponzi schemes in the recently deployed 10,000 smart contracts.
This highlights the severity of Ponzi contracts on Ethereum and the pressing need to effectively identify them.

\bibliographystyle{ACM-Reference-Format}
\bibliography{acmart}

\end{document}